\pdfoutput=1

\documentclass[12pt]{article}
\usepackage{graphicx}
\usepackage{amsmath}
\usepackage{amssymb}
\usepackage{amsfonts}
\usepackage{rotating}
\usepackage{overcite}
\usepackage{ccaption}
\usepackage{subfig}

\renewcommand{\baselinestretch}{1}   

\usepackage{amssymb,amsfonts,amsmath}

\begin{document}
\setcounter{page}{1}

\title{{\bf Extrinsic and intrinsic nucleosome positioning signals}}

\author{Alexandre V.~Morozov
\thanks{To whom correspondence should be addressed. E-mail: morozov@physics.rutgers.edu}~,
Karissa Fortney $^\dag$,
Daria A.~Gaykalova $^\ddag$, \\
Vasily M.~Studitsky $^\ddag$,
Jonathan Widom $^\dag$, and
Eric D. Siggia $^\star$ \\
{\footnotesize $^\ast$Department of Physics \& Astronomy and BioMaPS Institute for Quantitative Biology,} \\
{\footnotesize Rutgers University, 136 Frelinghuysen Road, Piscataway NJ 08854, U.S.A.} \\
{\footnotesize $^\dag$Department of Biochemistry, Molecular Biology, and Cell Biology,} \\
{\footnotesize Northwestern University, 2153 Sheridan Road, Evanston IL 60208, U.S.A.} \\
{\footnotesize $^\ddag$Department of Pharmacology, UMDNJ, Robert Wood Johnson Medical School,} \\
{\footnotesize 675 Hoes Lane, Piscataway NJ 08854, U.S.A.} \\
{\footnotesize $^\star$Center for Studies in Physics and Biology, The Rockefeller University,} \\
{\footnotesize 1230 York Avenue, New York NY 10065, U.S.A.}
}

\date{}
\maketitle


\begin{abstract}
In eukaryotic genomes, nucleosomes function to compact DNA and to regulate access to it both by simple
physical occlusion and by providing the substrate for numerous covalent epigenetic tags.
While nucleosome positions {\it in vitro} are determined by sequence alone, {\it in vivo} competition
with other DNA-binding factors and action of chromatin remodeling enzymes play a role that needs to be quantified.
We developed a biophysical model for the sequence dependence of DNA bending energies, and validated it
against a collection of {\it in vitro} free energies of nucleosome formation and a nucleosome crystal structure;
we also successfully designed both strong and poor histone binding sequences {\it ab initio}.
For {\it in vivo} data from {\it S.cerevisiae}, the strongest positioning signal came from
the competition with other factors.
Based on sequence alone, our model predicts that functional transcription factor binding sites have a tendency
to be covered by nucleosomes, but are uncovered {\it in vivo} because functional sites cluster within
a single nucleosome footprint, making transcription factors bind cooperatively.
Similarly a weak enhancement of nucleosome
binding in the TATA region for naked DNA becomes a strong depletion when the TATA-binding protein
is included, in quantitative agreement with experiment. Predictions at specific loci were also greatly
enhanced by including competing factors. Our physically grounded model distinguishes multiple ways in which genomic
sequence can influence nucleosome positions
and thus provides an alternative explanation for several important experimental findings.
\end{abstract}

\section{Introduction}

Genomic DNA is packaged into chromatin in eukaryotic cells.
The building block of chromatin is the nucleosome, \cite{Khorasanizadeh}
a 147 base pair (bp) DNA segment wrapped in $\sim 1.8$ superhelical coils
on the surface of a histone octamer. \cite{RichmondNCP}
The unstructured tails of the histones are the targets of numerous covalent
modifications \cite{Khorasanizadeh}
and may influence how the ordered nucleosome array folds into higher order chromatin structures.
Chromatin can both block access to DNA \cite{Boeger}
and juxtapose sites far apart on the linear sequence. \cite{Wallrath}

The regulation of gene expression by DNA-bound factors can be partially
understood from their binding energies and thermodynamics.
Nucleosomes formation energies vary by about 5 kcal/mol \textit{in vitro}, \cite{Thastrom}
comparable to the variation of DNA-binding energies in other factors. It is then natural to ask whether
the locations of nucleosomes \textit{in vivo} can be predicted thermodynamically,
either in isolation or in competition with other DNA-binding factors.
Under this scenario, the role of chromatin and its remodeling enzymes would be
catalytic, modifying the rate of assembly but not the final disposition of factors on DNA.
The alternative is that chromatin remodeling within a regulatory unit actively positions nucleosomes
to control access to the DNA, in analogy with motor proteins.

It has not been possible to quantify by genetics where living cells fall between these extremes.
Recent computational approaches \cite{Segal,Ioshikhes,Peckham}
have used collections of DNA positioning sequences isolated
\textit{in vivo} to train pattern matching tools that were then applied genome-wide.
However, the training data may not be representative of direct histone-DNA binding if other factors
reposition nucleosomes \textit{in vivo} by exclusion.
Furthermore, models based on alignments of nucleosome positioning sequences \cite{Segal,Ioshikhes}
require a choice of background or reference sequence and it is known that nucleotide
composition varies among functional categories of DNA.

\section{Results and Discussion}

\textbf{Biophysical model of a nucleosome core particle.}
For these reasons we developed a biophysical model for nucleosome formation that resolves the energy into
the sum of two potentials modeling the core histone-DNA attraction and the intra-DNA bending energy.
The first potential is assumed to be independent of the DNA sequence since there are few direct contacts
between the histone side chains and DNA bases. \cite{LugerNCP} For the DNA bending, we constructed an empirical
quadratic potential \cite{Olson_PNAS,Morozov} using a database of 101 non-homologous, non-histone
protein-DNA crystal structures.

More specifically, we model DNA base stacking energies by defining three displacements (rise, shift, and slide)
and three rotation angles (twist, roll and tilt) for each dinucleotide (Fig. 1a). \cite{Olson_PNAS}
Together the six degrees of freedom completely specify the spatial position of bp $i+1$ in the coordinate frame
of bp $i$ (Supporting Information (SI) Fig. 7).
Geometric transformations based on these degrees of freedom can be applied recursively to reconstruct
an arbitrary DNA conformation in Cartesian coordinates.
Conversely, the atomic coordinates in a crystal structure allow the inference
of the displacements and rotations for each dinucleotide (SI Methods).
The elastic force constants in the quadratic DNA bending energy are
inferred from the protein-DNA structural data by inverting a covariance matrix of deviations of dinucleotide geometric
parameters from their average values. \cite{Olson_PNAS} Strongly bent dinucleotides (with one or more geometric parameters
further than 3 standard deviations from the mean) are iteratively excluded from the data set (SI
Methods). Our model does not use any higher order
moments of empirical geometry distributions, which would lead to a non-quadratic
elastic potential; nor are there sufficient data to model more than successive base pairs.

We assume that the histone-DNA potential is at minimum along an ideal superhelix whose pitch and radius
are taken from the nucleosome crystal structure, \cite{RichmondNCP} and varies quadratically
when the DNA deviates from the superhelix. This sequence-independent term represents the balance
between the average attractive electrostatic (and other) interactions between the histones
and the DNA phosphate backbone, \cite{Arents} and steric exclusion between the proteins and DNA.

The sum of the bending and histone-DNA potentials is minimized to yield the elastic energy
and the dinucleotide geometries for each nucleosomal sequence (see Methods).
Because the total energy is quadratic, energy minimization is equivalent to solving a system of linear
equations and thus the algorithm can be applied on a genome-wide scale.
There is no background in this model, but the results may depend on the choice of the protein-DNA structural data set.
Since our bending energy is empirical and inferred from co-crystal structures, it lacks a physical energy scale.
By comparison with the worm-like chain model our units can be converted to kcal/mol through multiplication by 0.26
(SI Methods).
We position multiple nucleosomes and proteins bound to DNA using standard thermodynamics and enforce steric exclusion
between bound entities in any given configuration (SI Methods).
Our program DNABEND can compute the sequence-specific energy of DNA restrained to follow an arbitrary
spatial curve.

\textbf{Analysis of DNA geometries from the nucleosome crystal structure.}
One way to validate DNABEND is to predict the DNA conformation in the high-resolution
(1.9 \AA) nucleosome crystal structure \cite{RichmondNCP}
(PDB code 1kx5) using only the DNA sequence. The DNABEND predictions are significantly correlated with
the experimental geometries for twist, roll, tilt and slide ($r \ge 0.49$), but are less successful for shift and rise
(SI Fig. 10), although the peak positions are generally correct.
DNABEND does not reproduce rapid shift oscillations in the region between bps 35 and 105,
and underestimates the magnitude of observed deviations in rise and slide.
We under-predict slide since for certain key base pairs our training structures imply a much smaller mean value
of slide (\textit{e.g.} 0.18 \AA~ for CA steps) than that observed in the nucleosome structures (0.91 \AA~ for CA steps).
Changing just this one mean value gives more reasonable magnitudes of slide (SI Fig. 10, black curve).
Tolstorukov \textit{et al.} also observed that slide is large and positive in the nucleosome crystal structure,
and makes a significant contribution to the superhelical trajectory. \cite{Tolstorukov}
Because nucleosomal DNA is highly bent, different degrees of freedom are strongly coupled (SI Fig. 11a):
for example, base pairs tend to tilt and shift simultaneously to avoid a steric clash.
These couplings are much less pronounced in the non-histone protein-DNA complexes used to derive the elastic
energy model (SI Fig. 11b), but nonetheless appear prominently when the 1kx5 DNA is positioned by DNABEND
(SI Fig. 11c).

Analysis of available nucleosome crystal structures shows that the tight DNA wrapping is facilitated by sharp DNA kinks
if flexible dinucleotides (\textit{e.g.} 5'-CA/TG-3' or 5'-TA-3') are introduced
into the region where the minor groove faces the histone surface.
For other sequences and other structural regions the bending is distributed over several
dinucleotides. \cite{RichmondNCP}
We substituted all possible dinucleotides into the 1kx5 atomic structure (keeping DNA conformation fixed),
and computed the elastic energy for each sequence variant. The most sequence specific regions are those
where the minor groove faces the histone octamer (SI Fig. 8a).
The specificity is especially dramatic if DNA is kinked (\textit{e.g.} at positions 109, 121 and 131;
SI Fig. 10). \cite{RichmondNCP}
Although these positions are occupied by CA/TG dinucleotides in the crystal structure, the model assigns the lowest energy
to TA dinucleotides, consistent with the periodic TA signal previously observed in good nucleosome positioning
sequences \cite{Widom} (SI Fig. 8b).

\textbf{Comparison with {\it in vitro} measurements.}
DNABEND accurately predicts experimental free energies of nucleosome formation \cite{Thastrom,Shrader_I,Shrader_II}
(Fig. 2a, SI Fig. 12a).
In contrast, the alignment model of Segal \textit{et al.} \cite{Segal} trained on yeast nucleosomal
sequences has little predictive power for the sequences in this set, most of which were chemically synthesized.
DNABEND also correctly ranks sequences selected \textit{in vitro} for their ability to form stable
nucleosomes, or to be occupied by nucleosomes \textit{in vivo} (Fig. 2b). \cite{Lowary,Segal}
Because the alignment model is constructed using the latter set it
assigns anomalously low (more favorable) energies to some of the sequences from it,
and higher (less favorable) energies to the artificial sequences known
from experiment to have excellent binding affinities. \cite{Lowary,Thastrom}
Finally, DNABEND correctly ranks mouse genome sequences selected \textit{in vitro}
on the basis of their high or low binding affinity \cite{Kubista_I,Kubista_II} ($P = 1.42 \times 10^{-9}$),
whereas the alignment model has less resolution ($P = 1.73 \times 10^{-2}$), though on average it does assign better
scores to the high affinity set (SI Fig. 12b).

A further test of how DNABEND actually positions nucleosomes on DNA
can be provided by a collection of sequences
where \textit{in vitro} positions are known with 1-2 bp accuracy. We have determined nucleosome positions
on synthetic high-affinity sequences 601, 603, and 605 \cite{Lowary} using hydroxyl radical footprinting
(SI Figs. 13 and 14), and obtained 3 more sequences from the literature. \cite{Tolstorukov}
The measured position is always within 1-2 bp of a local minimum in our energy, and that energy minimum in 5 out of 6 cases
is within 0.5-1.0 kcal/mol of the global energy minimum
(SI Fig. 15; note that the total range of sequence-dependent binding energies is $\sim 5$ kcal/mol).

We also asked if DNABEND could be used to design DNA sequences with high and low
binding affinities for the histone octamer. Free energies of computationally designed sequences
were measured \textit{in vitro} using salt gradient dialysis \cite{Thastrom,Thastrom_II} (SI Table 1, SI Methods).
The free energy of the designed best sequence was lower than the free energy of the designed worst sequence, although only by
1.6 kcal/mol, which is less than the experimentally known range of free energies
(SI Table 1, SI Fig. 12a).
These results underscore both the ranking power and the limitations of our current DNA mechanics model.

\textbf{Periodic dinucleotide distributions in high and low energy sequences.}
DNABEND-selected nucleosome sequences exhibit periodic dinucleotide patterns that are consistent with those
determined experimentally: \cite{Lowary}
for example, with lowest energy sequences, 5'-AA/TT-3' and 5'-TA-3' dinucleotide frequencies
are highest in the negative roll regions (where the minor groove faces inward),
while 5'-GC-3' frequencies are shifted by $\sim 5$ bp (SI Fig. 16).
Surprisingly, the distributions of AT and AA/TT, TA dinucleotides are in phase,
despite a very low flexibility of the former (SI Fig. 8b).
It is possible that AT steps are used to flank a more flexible kinked dinucleotide.
We estimate the energy difference between the best and the worst 147 bp nucleosome forming
sequences to be 15.2 kcal/mol, with the energies of 95\% of genomic sequences separated by less than 6.4 kcal/mol
(SI Methods).
This is larger than the experimental range (SI Fig. 12a)
because nucleosomes cannot be forced in experiments to occupy the worst-possible location on a DNA, but instead
tend to find the most favorable locations with respect to the 10 bp helical twist.

\textbf{Prediction of nucleosome positions in the yeast genome.}
We use nucleosome energies (computed using the 147 bp superhelix)
and the binding energies of other regulatory factors
to construct a thermodynamic model in which nucleosomes form while competing with other proteins
for access to genomic sequence. A typical configuration thus contains multiple DNA-bound molecules of several types
and explicitly respects steric exclusion. We take all such configurations into account using dynamic
programming methods \cite{Durbin} that enable us to compute a Boltzmann-weighted statistical sum and thus the probability $P$ for each factor
to bind DNA starting at every possible position along the genome (Fig. 1b, SI Methods). \cite{Segal}
We also compute the occupancy of each genomic bp,
defined as its probability to be covered by any protein of a given type (Fig. 1b).
There are two adjustable parameters for each DNA-binding factor:
the temperature, which determines how many factors are bound with high probability, and the free protein concentration,
which determines the average occupancy.
With our choice of parameters (which have not been optimized, rather, were fixed to allow for comparison with previous studies
- SI Methods), the average nucleosome occupancy is 0.797, and stable,
non-overlapping nucleosomes (with $P > 0.5$) cover 16.3\% of the yeast genome.

\textbf{Bioinformatic models based on alignments of nucleosome positioning sequences.}
Several existing analyses of nucleosome positions are based on alignment models, and these in turn explicitly \cite{Segal}
or implicitly \cite{Ioshikhes} include a background model. We examined the sensitivity of one
alignment-based approach to the choice of background
by implementing it in two different ways: alignment model I is identical to Segal et al. \cite{Segal}
and uses genomic background for one strand and uniform background for the other,
while alignment model II employs genomic background
for both strands (SI Methods).
Comparing two models allows us to separate more robust predictions from those that depend
strongly on the implementation details.
Although the alignment models correlate well with each other ($r = 0.66$), we find a small \textit{negative}
correlation between DNABEND-predicted occupancies and energies
and those from the alignment model I (SI Fig. 18).
DNABEND energies are strongly correlated if two nucleosomes are separated by a multiple of 10 bp,
and anti-correlated if the nucleosomes are separated by a multiple of 5 bp, which puts the helical twist out of
phase (SI Fig. 17c).
Log scores predicted by the alignment model exhibit weaker oscillations (SI Fig. 17d),
probably due to ambiguities in aligning the training sequences.

Because all three models make somewhat different predictions of average nucleosome occupancies in broad genomic regions
(SI Fig. 19), additional experimental data are required to establish which model is the most accurate.
In contrast, all models predict that the distribution of center-to-center distances between
proximal stable nucleosomes (with $P \ge 0.5$) exhibits strong periodic oscillations (SI Fig. 20).
However, similar oscillations occur in a non-specific model where we first create a regular nucleosomal array and
then randomly label a given fraction of nucleosomes as stable (data not shown).
Because steric exclusion is respected by every model,
nucleosomes form regular arrays (SI Figs. 17a,b) which help induce the observed periodicity in the
positions of stable nucleosomes.

\textbf{Nucleosome depletion upstream of the ORFs.}
Microarray-based maps of \textit{in vivo} nucleosome positions show striking depletion of nucleosomes
from the promoter regions (Fig. 3a). \cite{Yuan,Ozsolak}
We find that this depletion is difficult to explain using nucleosome models alone.
However, the nucleosome-free region upstream of the open reading frames (ORFs) becomes much more
pronounced when TATA box-binding proteins (TBPs) and other factors
bind their cognate sites (Fig. 3a, SI Fig. 21).
Displaced nucleosomes are re-arranged in regular arrays on both sides of factor-occluded sites, \cite{Kornberg,Fedor} 
creating the characteristic oscillatory structure around the nucleosome-free region
which includes the occupancy peak over the translation start observed in both regular \cite{Yuan} and
H2A.Z \cite{Albert} nucleosomes (Fig. 3a and SI Fig. 22).
Similar oscillations occur when non-specific nucleosomes compete with TBP (SI Fig. 21),
making intrinsic nucleosome sequence preferences hard to disentangle from the larger phasing effects in this data set.

\textbf{Nucleosome occupancy of TATA boxes.}
Nonetheless, nucleosome stabilities can play a crucial role in gene activation: for example,
DNABEND predicts that both TATA boxes in the promoter of the yeast \textit{MEL1} ($\alpha$-galactosidase) gene
are occupied by a stable nucleosome, in agreement with the extremely low level of background gene expression
observed in \textit{MEL1} promoter-based reporter plasmids \cite{Ligr,Melcher} (Fig. 1b).
In contrast, the TATA elements of the \textit{CYC1} promoter were shown
to be intrinsically accessible \textit{in vivo}, \cite{Kuras,Chen}
resulting in high background expression levels. 
Consistent with these findings, we predict that one of the \textit{CYC1} TATA boxes has
intrinsically low nucleosome occupancy, and moreover that the nucleosome is easily displaced
in competition with TBP (Fig. 1b).

It has been suggested that TATA box-containing genes may be repressed through
steric exclusion of TBP by a nucleosome placed over the TATA box. \cite{Struhl}
In agreement with this hypothesis, 
in the absence of other factors DNABEND predicts slightly higher nucleosome occupancy over TATA boxes (Fig. 3b).
Stress-induced genes are TATA-rich and may be nucleosome-repressed under non-stress conditions. \cite{Struhl}
Thus, this prediction of DNABEND can be tested experimentally by measuring
nucleosome occupancy over TATA boxes of stress-induced genes in
non-inducing conditions, genome-wide (to ensure adequate statistics which
were not provided by the data sets available to us, \cite{Yuan} although it was found in a very recent
genome-wide study of nucleosome occupancy in yeast that promoters of stress-induced genes
tend to be covered by nucleosomes when cells are grown in YPD media, whereas
the transcription start site for the typical gene was nucleosome-free \cite{Lee}).

\textbf{Nucleosome-induced TF cooperativity and nucleosome depletion over TF binding sites.}
It has been known from chromatin immunoprecipitation experiments that some near-consensus DNA sequences
are not occupied by their cognate transcription factors (TFs), while poorer sites may be occupied. \cite{MacIsaac_I}
Thus it is natural to ask whether histone binding preferences can help distinguish functional and
non-functional TF binding sites. Segal \textit{et al.} found that nucleosome occupancy was intrinsically smaller
at functional sites \cite{Segal} (SI Fig. 24a), while DNABEND predicts the opposite (Fig. 4a, upper panel)
(but see contrary low resolution data from Liu \textit{et al.} \cite{Liu}).
However, if TFs are allowed to compete with nucleosomes at all sites, the nucleosomes become preferentially depleted
(and the TF occupancy becomes higher) at the functional sites (Fig. 4a, lower panel).
This depletion is not due to the slightly more favorable binding energies of the functional sites (SI Fig. 25a),
because it is reversed when their positions are randomized within intergenic regions (SI Fig. 24c).
Furthermore, DNABEND-predicted nucleosomes are not depleted over functional sites simply because
they are less stable - Fig. 4a showed the opposite. In fact,
functional sites tend to occur just upstream of the ORFs (SI Fig. 25b),
where DNABEND-predicted nucleosomes exhibit enhanced stability (SI Fig. 23).
The only remaining possibility is that the depletion of DNABEND-predicted nucleosomes is caused by the spatial clustering
of functional binding sites (Fig. 4b, SI Fig. 25c). In this scenario several DNA-binding
proteins cooperate to evict the nucleosome and thus enhance their own binding, in
a phenomenon known as the nucleosome-induced cooperativity (Fig. 4c). \cite{Miller,Adams}
The cooperativity does not depend on direct interactions between TFs and requires only that two or more TF sites
occur within a 147 bp nucleosomal footprint (Fig. 4d).
A model in which TFs compete with nucleosomes for their
cognate sites provides the best explanation for the strong nucleosome depletion over functional sites observed
in microarray experiments \cite{Yuan} (Fig. 4e).
Although it is not yet conventional to do so, \cite{MacIsaac_I}
it would be trivial to reward clustering within 147 bp in codes that predict regulatory sites.

\textbf{Predictions of experimentally mapped nucleosome positions.}
We used a set of 110 nucleosome positions reported in the literature for 13 genomic loci
to see how well DNABEND predictions match \textit{in vivo} chromatin structure (Fig. 5, SI Figs. 26a-n).
The predictive power of nucleosome models improves considerably when sequence-specific factors are
included (Fig. 5), though not in genome-wide sets (SI Fig. 27);
this result may be due to the lack of accurate genome-wide knowledge of TF binding sites
and energies.
For example, at $HML\alpha$, $HMR\mathbf{a}$, and recombination enhancer loci, nucleosomes are positioned in regular arrays
whose boundaries are determined by the origin recognition complex, Abf1, Rap1, and Mcm1/MAT$\alpha 2$
bound to their cognate sites (SI Figs. 26f-h). 

Both intrinsic sequence preferences and boundaries created by other factors contribute to positioning
nucleosomes:
on one hand, at the \textit{GAL1-10} locus a nucleosome covering a cluster of GAL4 sites is evicted,
making $\sim 200$ bp of promoter sequence accessible to TBP and other factors (SI Fig. 26a),
as observed \textit{in vivo}. \cite{Li}
On the other hand, GCN4 and ABF1 sites at the \textit{HIS3-PET56-DED1} locus are intrinsically nucleosome-depleted,
because histones have lower affinity for \textit{HIS3-PET56} and \textit{DED1} promoters
(Fig. 6, SI Fig. 26k). \cite{Sekinger}
DNABEND correctly predicts chromatin re-organization caused by sequence deletions
in the \textit{HIS3-PET56} promoter region: \cite{Sekinger} sequence-dependent nucleosome positions are
refined through the action of GCN4 and TBP to improve the agreement with experiment (SI Figs. 26l-n).

\section{Conclusion}

We have developed a DNA mechanics model capable of accurately predicting \textit{in vitro} free energies
of nucleosome formation, optimal base pairs in the minor and major grooves of nucleosomal DNA, and
DNA geometries specific to each base step.
We only get an agreement with the available genome scale data sets \cite{Yuan,Ozsolak} for nucleosome depletion
from the TATA box and functional TF binding sites when we include competition with the relevant factors.
In the absence of DNA-binding factors DNABEND predicts a weak enhancement of nucleosome occupancy over their sites and thus agrees
with the conjecture, not yet demonstrated on a genome-wide scale, that nucleosomes provide a default
repression. \cite{Struhl} The two alignment models we examined do not support this conjecture; it is conceivable that they
have implicitly captured a hybrid signal from both nucleosomes and DNA-bound factors.

The highest quality data on generic nucleosome occupancy come from the specific loci summarized in Fig. 5.
There is a relatively weak signal from all models in the absence of other factors, originating
from a few correct predictions at short distances.
Because DNABEND predicts correct binding energies, the weak signal suggests that nucleosomes positioned
by thermodynamics on bare DNA have relatively weak correlation with \textit{in vivo} positions, while
inclusion of other factors substantially improves the picture.
DNABEND presents a useful biophysical framework for analysis of \textit{in vivo} nucleosome
locations and TF-nucleosome competition.
It will be interesting to examine its predictions for metazoan genomes, and
to modulate gene expression levels in model systems through computational design of nucleosome occupancy profiles.

\section{Methods}

For full details, see SI Methods. DNABEND software and additional supporting data
(including a \textit{S.cerevisiae} nucleosome viewer)
are available on the Nucleosome Explorer website: \textit{http://nucleosome.rockefeller.edu}.

The total energy of a nucleosomal DNA is given by a weighted sum of two quadratic potentials:
\begin{equation} \label{Etot:main}
E = E_{el} + w E_{sh},
\end{equation}
where $E_{el}$ is the sequence-specific DNA elastic energy \cite{Olson_PNAS,Morozov}
and $E_{sh}$ is the histone-DNA interaction energy (see SI Methods).
The potentials are quadratic with respect to deviations
of global displacements $\delta \vec{d}_s$ and local angles
$\delta \vec{\Omega}_s$ from their ideal superhelical values.
The weight $w$ is fit to maximize the average correlation coefficient between
the distributions of geometric parameters observed in 1kx5 \cite{RichmondNCP}
and the DNABEND predictions (SI Fig. 10).
The conformation adopted by the DNA molecule is the one that minimizes its total energy $E$,
yielding a system of linear equations:
\begin{equation} \label{Ederiv:main}
\begin{array}{ccc}
{\partial E}/{\partial \delta \alpha^{s}_{i}} = 0 & s = 1 \dots N, & i = 1 \dots 6,
\end{array}
\end{equation}
where $\mathbf{\delta \alpha}^s = (\delta \vec{\Omega}_s, \delta \vec{d}_s)$ and $N$ is the number of dinucleotides.
The nucleosome energies (assumed to be given by $E_{el}$) and the energies of other DNA-binding factors
at each genomic position are then used as input to a dynamic programming algorithm \cite{Durbin}
which outputs binding probabilities and bp occupancies for each DNA element.

\section*{Acknowledgments}
A.V.M. was supported by the Lehman Brothers Foundation through a Leukemia and Lymphoma Society fellowship.
E.D.S. was funded by the NSF grant DMR-0129848.
J.W. acknowledges support from NIGMS grants R01 GM054692 and R01 GM058617,
and the use of instruments in the Keck Biophysics Facility at Northwestern University.
V.M.S. was supported by NSF (0549593) and NIH (R01 GM58650) grants.
We thank Eran Segal, Michael Y.~Tolstorukov, Wilma K.~Olson, and Victor B.~Zhurkin
for useful discussions and for sharing nucleosome data.

\newpage 


\newpage

\renewcommand{\topfraction}{.85}
\renewcommand{\bottomfraction}{.7}
\renewcommand{\textfraction}{.15}
\renewcommand{\floatpagefraction}{.66}
\renewcommand{\dbltopfraction}{.66}
\renewcommand{\dblfloatpagefraction}{.66}
\setcounter{topnumber}{9}
\setcounter{bottomnumber}{9}
\setcounter{totalnumber}{20}
\setcounter{dbltopnumber}{9}


\captionnamefont{\bfseries}
\captiontitlefont{\small\sffamily}
\renewcommand{\figurename}{Figure}
\captiondelim{: } 


\begin{figure}[tbph]
\centering
\includegraphics[scale=0.63]{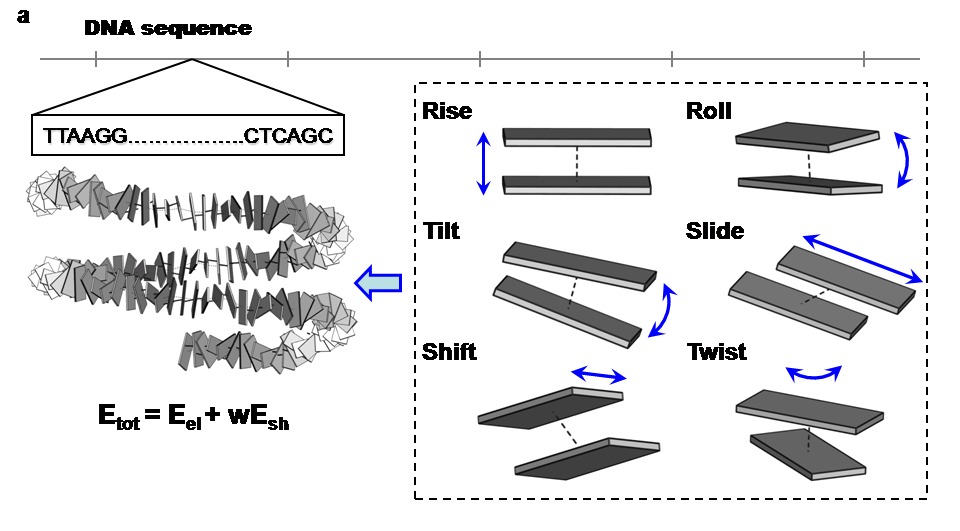}
\includegraphics[scale=0.63]{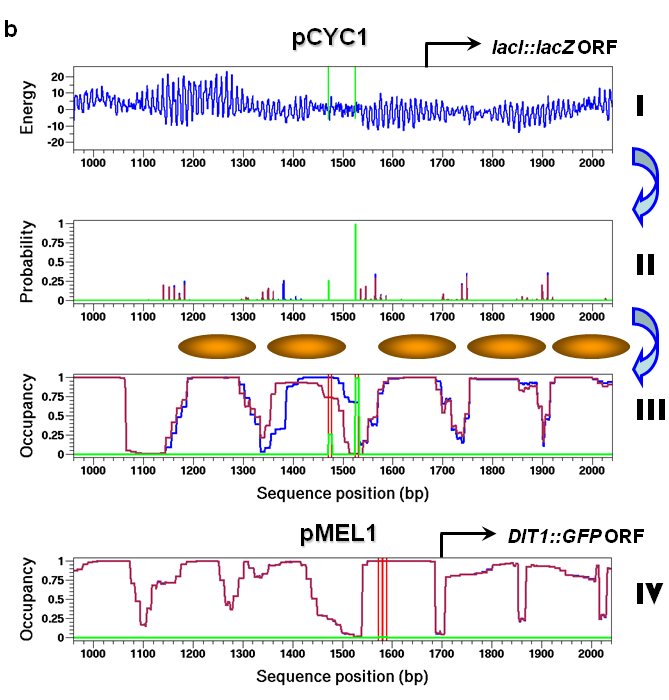}
\end{figure}

\begin{figure}[tbph]
\caption{
\textbf{DNA mechanics model of histone-DNA interactions is used to predict \textit{in vitro} and \textit{in vivo} nucleosome positions.}
a) Conformation of a single DNA base step (defined as two consecutive DNA base pairs in the $5' \to 3'$ direction)
is described by six geometric degrees of freedom: rise, shift, slide, twist, roll, and tilt. \cite{Olson_PNAS}
DNA base pairs are shown as rectangular blocks.
The nucleosome energy is a weighted sum of the elastic energy $E_{el}$ and
the restraint energy $E_{sh}$ which penalizes deviations of the DNA path from the ideal superhelix (see Methods).
b) Nucleosome positions explain background gene expression levels observed in reporter plasmids.
Panel I (from top). Blue: nucleosome energies (in arbitrary units)
in the $CYC1$ promoter region from the lacI::lacZ reporter plasmid, \cite{Chen}
with the 10-11 bp periodicity due to DNA helical twist.
Vertical green lines: energies of TBPs bound to two experimentally mapped TATA boxes \cite{Chen}
(other binding positions are not allowed).
Panel II. Probability of a nucleosome to start at each base pair, in the absence (blue)
and presence (maroon) of TBP. Some of the latter nucleosomes are also shown as orange ovals
(note that in general nucleosome positions may overlap).
Green: probability of a TBP to bind a TATA box.
Panel III. Nucleosome occupancy in the absence (blue) and presence (maroon) of TBP.
Green: TBP occupancy, red vertical lines: TATA box positions.
Arrows on the right correspond to the order of calculations.
Panel IV. Nucleosome occupancy of the \textit{MEL1} promoter region from the DIT1::GFP reporter vector \cite{Ligr}
(see \textit{CYC1} legend for the color scheme).
Note that blue and maroon occupancy profiles completely overlap.
}
\end{figure}

\begin{figure}[tbph]
\centering
\includegraphics[scale=0.90]{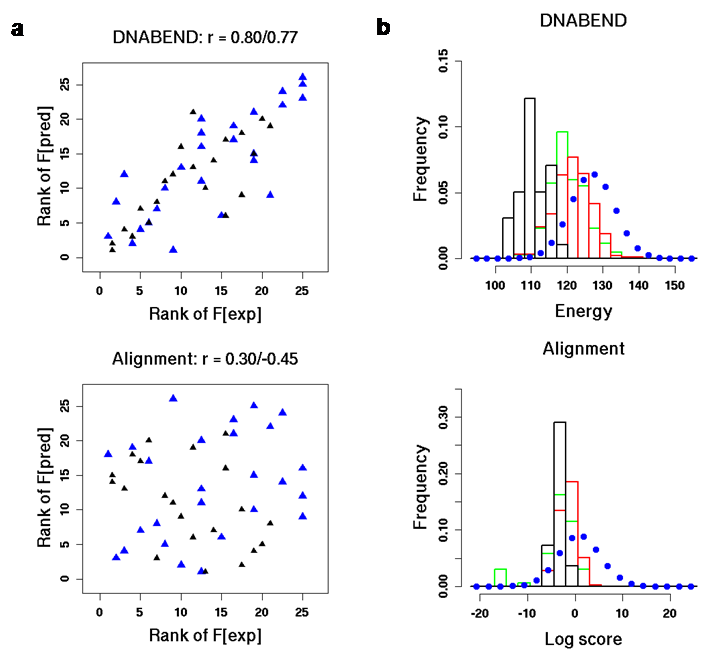}
\end{figure}

\begin{figure}[tbph]
\caption{
\textbf{DNABEND accurately ranks free energies of nucleosome formation and sets of nucleosome sequences.}
a) Nucleosome free energies of natural and
synthetic sequences ranked by DNABEND (upper panel) and by
a probabilistic model based on the alignment of sequences
extracted from yeast mononucleosomes (lower panel). \cite{Segal}
Blue triangles: Thastrom \textit{et al.} nucleosome dialysis assay, \cite{Thastrom}
black triangles: Shrader \textit{et al.} nucleosome exchange assay. \cite{Shrader_I,Shrader_II}
Free energies were computed using only the central 71 bp of the 147 bp nucleosomal site (cf. SI Fig. 12), because
competitive nucleosome reconstitution on DNAs with any lengths between 71 and 147 bp gives identical
apparent free energies, and quantitatively equivalent free energies are obtained
using either the full histone octamer or just the core histone tetramer. \cite{Thastrom_II,Segal}
b) Histograms of DNA elastic energies (in arbitrary units) and alignment model log scores
computed using the 147 bp nucleosomal site, consistent with sequence lengths found in the
\textit{in vitro} selection on the yeast genome. \cite{Segal}
Yeast genomic sequences are compared to three sets of sequences selected for
their nucleosome positioning ability. Blue: energies of all 147 bp long sequences from \textit{S.cerevisiae} chromosome III,
green: energies of sequences from a genome-wide \textit{in vivo} mononucleosome extraction assay \cite{Segal},
red: energies of sequences from an \textit{in vitro} selection assay on yeast genomic DNA \cite{Segal},
black: energies of sequences from a SELEX experiment on a large pool of chemically synthesized random DNA
molecules. \cite{Lowary} Sequences shorter than 147 bp were omitted from all selected sequence sets;
in sequences longer than 147 bp the most favorable energy was reported, taking both forward and reverse strands into account.
}
\end{figure}

\begin{figure}[tbph]
\centering
\includegraphics[scale=0.6]{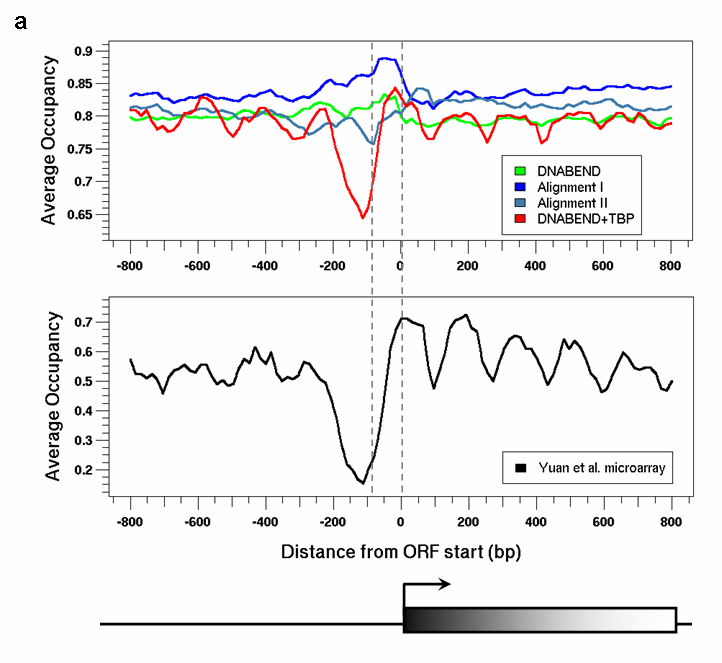}
\includegraphics[scale=0.6]{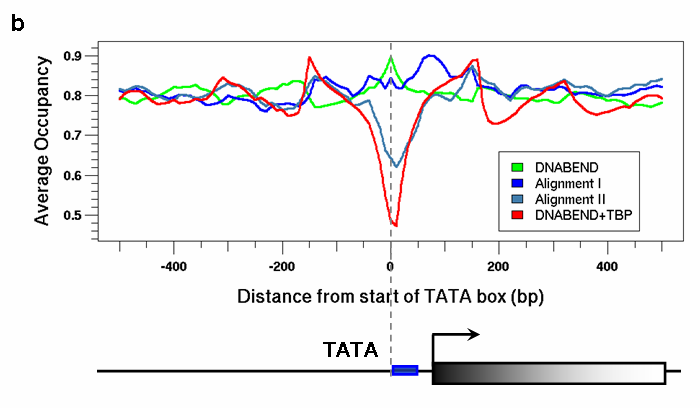}
\caption{
\textbf{Nucleosome-free regions are explained by bound TBPs.}
a) Average nucleosome occupancy plotted with respect to the translation start sites (black arrow) (a),
and TATA boxes \cite{Basehoar} (blue rectangle) (b).
Green: nucleosomes only (DNABEND), dark blue: nucleosomes only (alignment model I), \cite{Segal}
blue: nucleosomes only (alignment model II),
red: nucleosomes (DNABEND) competing with TBPs for the access to TATA boxes,
black: nucleosome positions inferred from a microarray measurement. \cite{Yuan}
}
\end{figure}

\begin{figure}[tbph]
\centering
\includegraphics[scale=0.68]{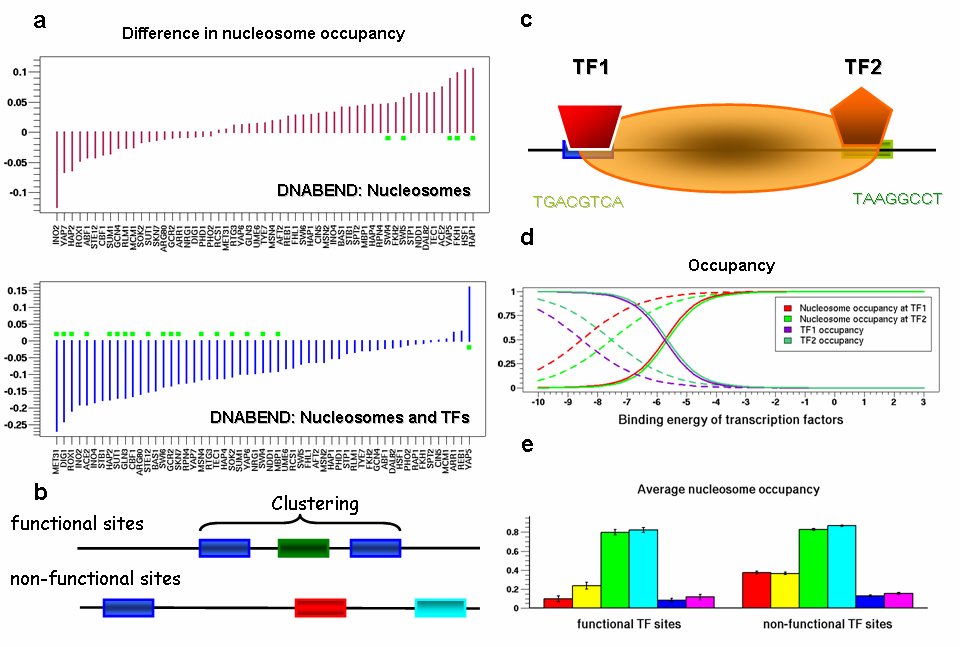}
\end{figure}

\begin{figure}[tbph]
\caption{
\textbf{Nucleosome-induced TF cooperativity explains nucleosome depletion over functional TF binding sites.}
a) Difference in the average nucleosome occupancy of functional and non-functional TF binding sites. \cite{MacIsaac_I}
The difference is negative if the functional sites are depleted of nucleosomes.
Green squares indicate statistically significant occupancy differences ($P < 0.05$).
Upper panel: nucleosomes only, lower panel: nucleosomes competing with TFs (allowed to bind both functional and
non-functional sites).
Functional sites are conserved in at least 3 out of 4 \textit{sensu stricto} yeast species
and are strongly supported by ChIP-chip data (probes bound by the factor with $P < 0.001$),
whereas the non-functional sites are not conserved and have weaker evidence for \textit{in vivo}
binding ($P < 0.005$). \cite{MacIsaac_I}
b) Functional binding sites are clustered (SI Fig. 25c) and thus exhibit nucleosome-induced cooperativity. \cite{Adams,Miller}
c) Nucleosome-induced cooperativity requires that a single nucleosome overlap two or more TF binding sites.
d) Nucleosome-induced cooperativity in a model system with two TF sites covered by a single nucleosome.
Nucleosome and TF occupancies at TF sites are plotted as a function of the binding energy (assumed to be the
same for both factors).
Dashed curves: occupancies in the absence of the other TF (no cooperativity),
solid curves: occupancies in the presence of both TFs (cooperativity).
e) Average nucleosome occupancy of functional and non-functional TF binding sites on chromosome III.
Red: Yuan \textit{et al.} measurements, \cite{Yuan}
yellow: nucleosomes in the presence of TFs (DNABEND), green/blue: nucleosomes in the absence of TFs
(DNABEND/alignment model I \cite{Segal}),
dark blue/violet: stable nucleosomes in the absence of TFs ($P \ge 0.5$, DNABEND/alignment model I).
The best discrimination is achieved by DNABEND when TFs are included.
}
\end{figure}

\begin{figure}[tbph]
\centering
\includegraphics[scale=0.80]{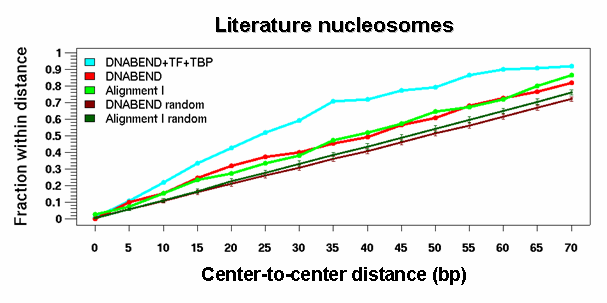}
\caption{
\textbf{Both TF and histone binding must be taken into account for accurate prediction of \textit{in vivo} nucleosome positions.}
Fraction of correctly predicted nucleosomes as a function of the center-to-center distance between predicted
nucleosomes and 110 nucleosomes with positions known from the literature. \cite{Segal,Sekinger}
In the absence of sequence-specific proteins, the fraction of correctly predicted nucleosomes is $\ge 0.5$ only
for the \textit{PHO5}, $HMR\mathbf{a}$, and \textit{HIS3} deletion loci with DNABEND (SI Fig. 26),
and for the \textit{GAL1-10}, \textit{STE6} loci with alignment model I, \cite{Segal} for distances $\le 20$ bp.
See SI Fig. 27 for details of implementation, including the definition of the null (random) model
with steric exclusion.
Results from the alignment model II are similar (data not shown).
}
\end{figure}

\begin{figure}[tbph]
\centering
\includegraphics[scale=0.80]{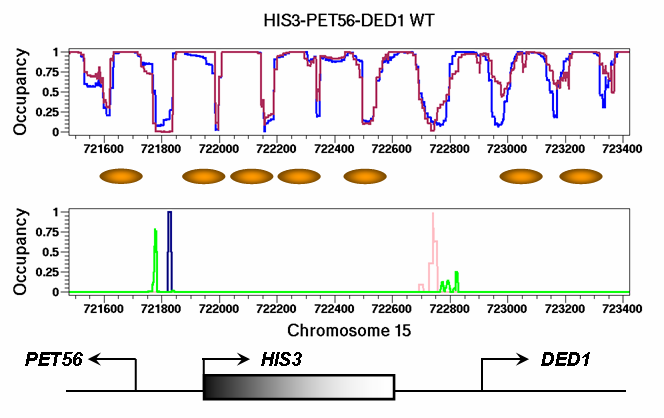}
\caption{
\textbf{Detailed view of the \textit{HIS3-PET56-DED1} wildtype locus.}
Upper panel: DNABEND-predicted nucleosome occupancy,
both with (maroon) and without (blue) other DNA-binding factors.
Lower panel: TBP occupancy (green), GCN4 occupancy (dark blue), and ABF1 occupancy (pink).
Orange ovals: experimental nucleosome positions mapped by Sekinger \textit{et al.} \cite{Sekinger}
Black arrows indicate translation start sites.
We used a GCN4 weight matrix from Morozov \textit{et al.} \cite{Morozov_II} and an ABF1 weight matrix
from MacIsaac \textit{et al.} \cite{MacIsaac_I} to compute TF DNA-binding energies.
The TBP weight matrix was derived using an alignment of TATA box sites from Basehoar \textit{et al.} \cite{Basehoar}
See SI Fig. 26 for further details.
Note that the region around bp 721800 is intrinsically devoid of nucleosomes, \cite{Sekinger}
allowing TFs to bind in the nucleosome-free gap.
}
\end{figure}

\clearpage


\begin{center}
\Large{\textbf{Supporting Information for: \\
Extrinsic and Intrinsic Nucleosome Positioning Signals}}
\end{center}

\section{Supplementary Methods}

\subsection{DNA geometry}

We model each DNA basepair as a rigid body and specify its position in space using a local coordinate frame attached to
each basepair and defined using
$R_{i} = [ \vec{n}_x^i, \vec{n}_y^i, \vec{n}_z^i ]$ and $\vec{r}_i$ ($i = 0 \dots N$). Here
$\vec{n}_{x,y,z}^i$ are the orthonormal basis vectors of the local frame, and $\vec{r}_i$
is its origin in the fixed global coordinate frame.
Both the basis vectors and the vector to the origin can be expressed in terms of
6 independent parameters that are sufficient for reconstructing the spatial position
of any rigid body. \cite{SI_Goldstein}
Thus apart from a single global translation and rotation an arbitrary DNA conformation with $N+1$ basepairs
is uniquely specified if $N$ sets of 6 geometric parameters are known. By convention, \cite{SI_Lu_3DNA,SI_SchnapsI,SI_SchnapsII,SI_ElHassan}
the geometric parameters are chosen to be the three angular degrees of freedom which define unit vectors
of the local frame attached to the basepair $i$ in the local frame attached to the basepair $i-1$,
and the displacement vector which gives the origin of frame $i$ with respect to origin of frame $i-1$:
$\alpha_i = (\vec{\Omega}_i, \vec{d}_i), ~~i = 1 \dots N$.
Here $\vec{\Omega}_i = \{\Omega_i, \rho_i, t_i\}$ are the helical twist, roll and tilt angles,
and $\vec{d}_i$ is the displacement vector with the x,y,z components
called slide, shift and rise (Fig. 1a). \cite{SI_Lu_3DNA,SI_ElHassan,SI_SchnapsI,SI_SchnapsII}
Since the geometric parameters which specify the spatial position of the basepair $i$ are defined with respect
to the frame rigidly attached to the previous basepair, their values capture local deviations in the DNA conformation.
However, they can also be used to recursively construct the rotation matrix for the basepair $i$ in the global frame:
\begin{equation} \label{Ri}
R_{i} = R_{i-1} T_i ~~(i = 1 \dots N),
\end{equation}
where each $T_i$ matrix is the product of three rotations:

\begin{equation} \label{Ti}
T_i = \mathbb{R}_z (-\frac{\Omega_i}{2} + \phi_i) \mathbb{R}_y (\Gamma_i) \mathbb{R}_z (-\frac{\Omega_i}{2} - \phi_i).
\end{equation}

Here,
\begin{equation*} \label{singleRy}
\mathbb{R}_y (\theta) = \left( \begin{array}{ccc}
\cos \theta & 0 & \sin \theta \\
0 & 1 & 0 \\
-\sin \theta & 0 & \cos \theta
\end{array} \right),
\end{equation*}

\begin{equation*} \label{singleRz}
\mathbb{R}_z (\theta) = \left( \begin{array}{ccc}
\cos \theta & \sin \theta & 0 \\
-\sin \theta & \cos \theta & 0 \\
0 & 0 & 1
\end{array} \right),
\end{equation*}

\begin{equation*} \label{GammaPhi}
\left\{ \begin{array}{l}
\Gamma_i = (\rho^2_i + t^2_i)^{1/2}, \\
\cos \phi_i = \rho_i/\Gamma_i, \\
\sin \phi_i = t_i/\Gamma_i.
\end{array} \right.
\end{equation*}

Note that $\mathbb{R}_y (\theta)$ and $\mathbb{R}_z (\theta)$ are the rotation matrices around the y and z axes, respectively.
The middle term on the right-hand side of Equation (\ref{Ti}) introduces both roll and tilt with a single rotation
through angle $\Gamma_i$ around a roll-tilt axis.
The roll-tilt axis lies in the x-y plane of the mid-step coordinate frame (mid-step triad, or MST: Supplementary Fig. 7),
and is in general inclined at an angle $\phi_i$ to the y-axis. \cite{SI_ElHassan}
In this scheme, it is the mid-step triad that is used to transform the displacement vector $\vec{d}_i$ into the fixed global frame:
$\vec{d}_i^{(g)} = R_{i}^{MST} \vec{d}_i$.
Introducing a single roll-tilt rotation axis is preferable to separate roll and tilt
rotations because it eliminates the non-commutativity problem associated with the roll and tilt
rotation operations. \cite{SI_ElHassan,SI_SchnapsI}

Similarly to the basepair rotation matrices, the mid-step triads are also constructed recursively:
\begin{equation} \label{Ri:MST}
R_{i}^{MST} = R_{i-1} T_i^{MST},
\end{equation}
where

\begin{equation} \label{Tmst}
T_i^{MST} = \mathbb{R}_z (-\frac{\Omega_i}{2} + \phi_i) \mathbb{R}_y (\frac{\Gamma_i}{2}) \mathbb{R}_z (- \phi_i).
\end{equation}
Thus having a complete set of local geometric parameters $\alpha_i$ is equivalent to knowing the global DNA conformation:
first, the recursive relation (\ref{Ri}) is employed to determine the orientations of all basepair coordinate frames
(except for the first one which has to be fixed in space independently and thus provides the overall orientation and position
of the DNA molecule). Second, Equation (\ref{Ri:MST}) is used to construct the MST frames, which are then employed to 
transform all local displacements $\vec{d}_i$ into the global frame. Finally, all displacement vectors are added up vectorially to
determine the origins of the basepair coordinate frames. If necessary, the basepair frames can then be used to
reconstruct the positions of all DNA basepair atoms (using an idealized representation which neglects flexibility
of bases in a basepair). Note that knowing a basepair coordinate frame is in general insufficient for
predicting the positions of the phosphate backbone and sugar ring atoms because of their additional degrees of freedom.
Finally we remark without showing the details of the calculations that the inverse problem is also well-defined:
a full set of basepair and MST rotation matrices in the global frame is sufficient
to reconstruct all local degrees of freedom $\{ \alpha_i \}$.
We have implemented the solution to the inverse problem following a previously published description. \cite{SI_SchnapsI}

While the helical twist serves to rotate the consecutive DNA basepairs with respect to one another,
introducing non-zero roll and tilt angles imposes curvature onto the DNA conformation.
Indeed, if both roll and tilt are negligible $T_i$ is simply a rotation through $-\Omega$
around the z-axis (which coincides with the helical axis in this case), and $T_i^{MST}$
is a rotation through $-\Omega/2$ around the same axis. The action of roll and tilt
can be seen more clearly by considering the curvature vector, defined as the difference between
the tangent vectors for two consecutive basepairs: $\vec{\kappa}_i = \vec{n}_z^{i+1} - \vec{n}_z^{i}$.
The expression for curvature is simplified in the limit of small roll and tilt:
since the magnitude of roll ($< 20^{\circ}$) and tilt ($< 10^{\circ}$) are typically much smaller than the helical twist
per basepair ($\simeq 36^{\circ}$), $T_{i}$ can be expanded to ${\cal O}(\rho, t)$:

\begin{equation} \label{Ti_linear}
T_i(\Omega,\rho,t) = \left( \begin{array}{rrr}
\cos \Omega & -\sin \Omega & \rho \cos (\Omega/2) + t \sin (\Omega/2) \\
\sin \Omega & \cos \Omega & \rho \sin (\Omega/2) - t \cos (\Omega/2) \\
- \rho \cos (\Omega/2) + t \sin (\Omega/2) & \rho \sin (\Omega/2) + t \cos (\Omega/2) & 1
\end{array} \right),
\end{equation}


Because we are primarily interested in the magnitude of the curvature vector,
we can compute it in the local frame of basepair $i$:
$\vec{\kappa}_i = T_i \widehat{z} - \widehat{z},$ where $\widehat{z} = (0, 0, 1)$ is the z-axis unit vector.
Using Equation (\ref{Ti_linear}) we obtain:
\begin{equation*}
\vec{\kappa}_i = \left( \begin{array}{c}
\rho_i \cos (\Omega_i/2) + t_i \sin (\Omega_i/2) \\
\rho_i \sin (\Omega_i/2) - t_i \cos (\Omega_i/2) \\
0
\end{array} \right)
\end{equation*}
Thus in this limit the magnitude of the curvature vector contains equal contributions from both roll and tilt:
$|\vec{\kappa}_i| = (\rho_i^2 + t_i^2)^{1/2} = \Gamma_i$.
Furthermore, the roll and tilt contributions to curvature are shifted by $90^{\circ}$ with respect to one another.

Transformations between local and global descriptions of the DNA molecule are important
because the sequence-specific DNA elastic energy is best described in terms of the local degrees of freedom $\{ \alpha_i \}$,
whereas it is the global coordinates that can be most naturally restrained to follow an arbitrary spatial curve.

\subsection{Ideal superhelix}

In nucleosome core particles the spatial curve used to restrain DNA conformation is an ideal left-handed superhelix
with pitch $P$ and radius $R$, described by the following parametric equation:
\begin{equation} \label{ideal:t}
\vec{\mathbf r} (t) = \left\{ \begin{array}{l}
R \cos ({2 \pi t}/{P}), \\
R \sin ({2 \pi t}/{P}), \\
-t.
\end{array} \right.
\end{equation}

Using the arc length $s = \sqrt{(2 \pi R / P)^2 + 1} ~t$ to parameterize the curve we obtain:
\begin{equation} \label{ideal:s}
\vec{\mathbf r} (s) = \left\{ \begin{array}{r}
R \cos (s/R_{eff}), \\
R \sin (s/R_{eff}), \\
-(P/2 \pi R_{eff}) s
\end{array} \right.
\end{equation}
where $R_{eff} = \sqrt{R^2 + (P/2 \pi)^2}$.
A local frame at position $s$ is given by a set of three orthonormal Frenet vectors (tangent, normal, and binormal):
\begin{equation} \label{tnb}
\left\{ \begin{array}{lll}
\vec{\mathbf t} (s) &=& d \vec{\mathbf r} / ds, \\
\vec{\mathbf n} (s) &=& d \vec{\mathbf t} / ds / |d \vec{\mathbf t} / ds|, \\
\vec{\mathbf b} (s) &=& \vec{\mathbf t} \times \vec{\mathbf n}.
\end{array} \right.
\end{equation}

We position $N+1$ sets of Frenet basis vectors equidistantly along the superhelical curve:
$s_{i} = 2 \pi R_{eff} \alpha (i / N)$ ($i = 0 \dots N$), where $\alpha$ is the number of nucleosomal superhelical
turns. Fitting an ideal superhelix to the high-resolution crystal structure of the nucleosome core particle \cite{SI_RichmondNCP}
reveals that
\begin{equation} \label{ideal:params}
\left\{ \begin{array}{l}
\alpha = 1.84, \\
P = 25.9 ~{\text \AA}, \\
R = 41.9 ~{\text \AA}
\end{array} \right.
\end{equation}

In order to transform the slowly rotating Frenet basis vectors into the helically twisted local frames
associated with DNA basepairs, we impose an additional helical rotation around the tangent vector $\vec{\mathbf t}$:
\begin{equation} \label{nb:rot}
\left\{ \begin{array}{lll}
\vec{\mathbf n}' (s_i) &=& \mathbb{R}_z (\Omega_{tot}^{i}) \vec{\mathbf n} (s_i), \\
\vec{\mathbf b}' (s_i) &=& \mathbb{R}_z (\Omega_{tot}^{i}) \vec{\mathbf b} (s_i).
\end{array} \right.
\end{equation}
Here $\Omega_{tot}^{i} = i \Omega_{0}$ is the cumulative helical twist, $\Omega_{0} = 34.696^{\circ}$ is the average helical twist
from the structure, and the rotation specified by Equations (\ref{nb:rot}) is performed in the
$(\vec{\mathbf t}, \vec{\mathbf n}, \vec{\mathbf b})$ basis. Similarly, the MST frames are located at 
$s_{i} = 2 \pi R_{eff} \alpha (i - 1/2)/ N$ ($i = 1 \dots N$), and are rotated through $(i - 1/2) \Omega_{0}$
to bring them into register with the helical twist.

A superhelix described by Equation (\ref{ideal:t}) has constant curvature which is manifested by the constant difference
between the consecutive tangent vectors: $|\vec{\mathbf t} (s_{i+1}) - \vec{\mathbf t} (s_{i})| = 2 \sin ({\pi \alpha}/{N})$.
This allows us to determine the length of the roll-tilt vector: $\Gamma_i = 4.53^{\circ}, \forall i$
(which also gives the maximum value of roll and tilt).
In this idealized picture, roll and tilt can be assumed to make equal contributions to the curvature. Indeed, reconstruction of the local
geometric parameters $\{ \alpha \}$ from the full set of helically twisted Frenet frames results in
$\vec{\Omega}_i = (\Omega_0, \Gamma_i \cos(\Omega_{tot}^{i} + \phi_0), \Gamma_i \sin(\Omega_{tot}^{i} + \phi_0))$
and $\vec{d}_i = (0, 0, d)$, where $d = 3.333$~\AA~ and $\phi_0$ is the initial phase determined by the
location of the first basepair.
Thus twist and rise are constant for every basepair in the ideal superhelix, slide and shift
are zero, whereas roll and tilt exhibit oscillations resulting from the superhelical curvature,
and shifted by $90^{\circ}$ with respect to one another (Supplementary Fig. 10).

\subsection{DNA conformational energy}

The total DNA conformation energy is a weighted sum of two terms:
the sequence-specific DNA elastic energy designed to penalize deviations of the local geometric parameters from their average values
observed in the protein-DNA structural database, and the restraint energy which penalizes deviations of the DNA molecule
from the nucleosomal superhelix. Thus all interactions (primarily of electrostatic nature)
between the DNA molecule and the histone octamer are modeled implicitly by imposing the superhelical restraint.

\vspace{0.2in}
\noindent \textbf{DNA elastic energy.}
Using the local degrees of freedom $\alpha_i$,
we represent the sequence-specific DNA elastic energy by a quadratic potential: \cite{SI_Olson_PNAS}
\begin{equation} \label{E_WO}
E_{el} = \frac{1}{2} \sum_{s=1}^{N} [\alpha^{s} - \langle \alpha^{n(s)} \rangle]^T F^{n(s)} [\alpha^{s} - \langle \alpha^{n(s)} \rangle],
\end{equation}
where the sum runs over all consecutive dinucleotides (basesteps) $s$ and
$\langle \alpha^{n} \rangle$ are the average values computed for the basestep type $n = {AA, AC, AG, .. ,TT}$
using a collection of oligonucleotides extracted from a set of 101 non-homologous protein-DNA structures. \cite{SI_Morozov}
The matrix of force constants $F^{n}$ is evaluated by inverting the covariance matrix $C^{n}$ of deviations of
local geometric parameters from their average values (${\alpha^{n} - \langle \alpha^{n} \rangle}$): \cite{SI_Olson_PNAS}
\begin{equation} \label{f:inv}
\begin{array}{cc}
(F^{n})^{-1} = C^n, & 
\mathrm{where} ~~~C^{n}_{ij} = \langle (\alpha^{n}_i - \langle \alpha^{n}_i \rangle) (\alpha^{n}_j - \langle \alpha^{n}_j \rangle) \rangle.
\end{array}
\end{equation}

Note that our elastic energy model utilizes only and first and second moments of the empirical geometry
distributions, and thus disregards all non-Gaussian effects such as skewness and bimodality. Including higher order
moments amounts to introducing non-quadratic corrections to the elastic energy model, which makes the computation
much less tractable. This problem is partially alleviated however by removing dinucleotides
from the data set if one or more of their geometric parameters are further than
3 standard deviations away from the mean. The mean is then recomputed
and the procedure is repeated until convergence. \cite{SI_Olson_PNAS}
DNA elastic energy is a function of 10 basestep type - dependent average values for each of the 6 local degrees of freedom
(all averages and covariances for the dinucleotides related by reverse complementarity are equal by construction), and of
15 independent force constants in each symmetric 6D matrix $F^n ~~(n = 1 \dots 10)$.

In order to be able to combine the DNA elastic term with the superhelical restraint term defined in the global frame,
we apply the following coordinate transformation to Equation (\ref{E_WO}):

\begin{equation} \label{Rs}
{\mathbf R}_s = \left( \begin{array}{cc}
\mathbf{1} & \mathbf{0} \\
\mathbf{0} & R_s^{MST} \\
\end{array} \right)
\end{equation}
($\mathbf{1}$ and $\mathbf{0}$ denote 3D unit and zero matrices, respectively). Applying ${\mathbf R}_s$ to the local
degrees of freedom leaves the angles (twist, roll, and tilt) invariant, but transforms the local displacements 
(shift, slide, and rise) into the global frame.
To keep the elastic energy invariant, the force constants have to be transformed as well:
$\mathbf{F}^n = {\mathbf R}_s F^n {\mathbf R}_s^{-1}$. Equation (\ref{E_WO}) then becomes:

\begin{equation} \label{E_WO_global}
E_{el} = \frac{1}{2} \sum_{s=1}^{N} [{\mathbf R}_s (\alpha^{s} - \langle \alpha^{n(s)} \rangle) ]^T \mathbf{F}^{n(s)}
[{\mathbf R}_s (\alpha^{s} - \langle \alpha^{n(s)} \rangle) ],
\end{equation}

Finally, it is convenient to change variables so that all degrees of freedom are expressed in terms of their deviations
from the ideal superhelix:
\begin{equation} \label{DoF:dev}
\left\{ \begin{array}{l}
\vec{d}_s^{(g)} = \vec{d}_s^{(g),0} + \delta \vec{d}_s, \\
\vec{\Omega}_s = \vec{\Omega}_s^{0} + \delta \vec{\Omega}_s.
\end{array} \right.
\end{equation}

Note that all the transformations described in this section involve no additional approximations to the original DNA elastic
energy (Equation (\ref{E_WO})). Thus we are free to choose the most convenient rotation matrix in Equation (\ref{Rs}),
and use $R_s^{MST} |_0$ from the ideal superhelix in Equation (\ref{E_WO_global}).

\vspace{0.2in}
\noindent \textbf{Superhelical restraint energy.}
Superhelical restraint energy is used to bend nucleosomal DNA into the superhelical shape. We define the restraint energy as
the quadratic potential:
\begin{equation} \label{Econstr}
E_{sh} = \sum_{s=1}^{N} (\vec{r}_s - \vec{r}_s^{0})^2,
\end{equation}
where $\vec{r}_s$ and $\vec{r}_s^{0}$ are the nucleosomal and the ideal superhelix radius-vectors in the global frame
($s = 1 \dots N$):

\begin{equation} \label{rad_vec}
\left\{ \begin{array}{l}
\vec{r}_s = \vec{r}_0 + \sum_{j=1}^s \vec{d}_j^{(g)}, \\
\vec{r}_s^{0} = \vec{r}_0^{0} + \sum_{j=1}^s \vec{d}_j^{(g),0}
\end{array} \right.
\end{equation}

Then to the lowest order the difference between the radius vectors is given by:

\begin{equation} \label{d:exp}
r_{s}^{\beta} - r_{s}^{0,\beta} =
\sum_{j=1}^s \left[
\delta d_{j}^{\beta}
+ \sum_{\alpha=1}^{3} \sum_{j'=1}^j b^{\alpha \beta}_{jj'} \delta \Omega_{j'}^{\alpha} \right],
\end{equation}
where $\alpha,\beta = 1 .. 3$ label the vector components, and
\begin{equation*}
b^{\alpha \beta}_{jj'} = \left( \frac{\partial R_j^{MST}}{\partial \Omega_{j'}^{\alpha}} |_0
\vec{d}_j^0 \right)^{\beta}
\end{equation*}
are the connectivity coefficients constructed as a product of the first derivative of the rotation matrix
with respect to the frame angles $\Omega_{j'}^{\alpha}$ and the ideal displacements $\vec{d}_j^0$ in the local frame.
The first term in Equation (\ref{d:exp}) represents the net change in the global radius vector $r_{s}^{\beta}$ caused by the changes
in the displacements $\vec{d}_{j}$ up to and including the basestep $s$,
while the second term reflects the change in the global radius vector resulting
from modifying one of the rotation angles $\Omega_{j}^{\alpha}, ~j \le s$.
Note that changing a single rotation angle at position $j$ affects every downstream basepair position linearly by introducing a kink
into the DNA chain, whereas making a displacement change at that position simply shifts all downstream coordinates by a constant amount.
The first derivative of the rotation matrix is evaluated as:
\begin{equation*}
\frac{\partial R_j^{MST}}{\partial \Omega_{j'}^{\alpha}} |_0 = \left\{ \begin{array}{ll}
T_0 \dots T_{j-1} \frac{\partial T_j^{MST}}{\partial \Omega_{j}^{\alpha}} & j' = j \\
T_0 \dots T_{j'-1} \frac{\partial T_{j'}}{\partial \Omega_{j'}^{\alpha}} T_{j'+1} \dots T_{j-1} T_j^{MST} &  1 \le j' < j
\end{array} \right.
\end{equation*}

Upon substitution of the expansion (\ref{d:exp}) into the restraint energy we obtain an effective quadratic potential:
\begin{equation} \label{Econstr:quad}
E_{sh} = \sum_{i,j=1}^{N} {\mathbf{\delta \alpha}^i}^T \mathbf{G}^{ij} \mathbf{\delta \alpha}^j,
\end{equation}
where $\mathbf{\delta \alpha}^i = (\delta \vec{\Omega}_i, \delta \vec{d}_i)$, and the 6x6 matrix of force constants is given
by three distinct 3x3 submatrices:
\begin{equation*}
\mathbf{G}^{ij} = \left( \begin{array}{cc}
H_{ij} & F_{ij} \\
F_{ij} & G_{ij}
\end{array} \right).
\end{equation*}

Here,
\begin{eqnarray*}
G_{ij}^{\alpha\beta} &=& \delta_{\alpha\beta} \sum_{s=1}^N \theta_{si} \theta_{sj}, \\
F_{ij}^{\alpha\beta} &=& \sum_{s,l=1}^N \theta_{si} \theta_{sj} \theta_{lj} b_{lj}^{\beta\alpha}, \\
H_{ij}^{\alpha\beta} &=& \sum_{s,k,l=1}^N \theta_{sk} \theta_{sl} \theta_{ki} \theta_{lj} \sum_{\gamma=1}^3
b_{ki}^{\alpha\gamma} b_{lj}^{\beta\gamma},
\end{eqnarray*}
where
\begin{equation*}
\theta_{ij} = \left\{ \begin{array}{cc}
1, & i \ge j \\
0, & i < j
\end{array} \right.
\end{equation*}
is the theta function, and $\delta_{\alpha\beta}$ is the Kronecker delta.
$G_{ij}^{\alpha\beta}$ couples displacements with displacements, $F_{ij}^{\alpha\beta}$
couples displacements with angles (and thus has one connectivity coefficient), and $H_{ij}^{\alpha\beta}$
couples angles with angles through two connectivity coefficients. Summing over some of the theta functions we obtain:

\begin{eqnarray*}
G_{ij}^{\alpha\beta} &=& \delta_{\alpha\beta} B(i,j), \\
F_{ij}^{\alpha\beta} &=& \sum_{l=j}^{N} B(i,l) b_{lj}^{\alpha\beta}, \\
H_{ij}^{\alpha\beta} &=& \sum_{k=i}^{N} \sum_{l=j}^{N} B(k,l) \sum_{\gamma=1}^{3} b_{ki}^{\alpha\gamma} b_{lj}^{\beta\gamma},
\end{eqnarray*}
where $B(i,j) = \sum_{s=1}^N \theta_{si} \theta_{sj} = N + 1 - \max(i,j)$.

\vspace{0.2in}
\noindent \textbf{Total energy and DNA conformation minimization.}
The total energy of nucleosomal DNA is given by:
\begin{equation} \label{Etot}
E = E_{el} + w E_{sh},
\end{equation}
where $w$ is the fitting weight introduced to capture the balance
between favorable histone-DNA interactions and the unfavorable energy of bending DNA into the nucleosomal superhelix.
We fit $w$ to maximize the average correlation coefficient between
the distributions of geometric parameters observed in the high-resolution crystal structure of the nucleosome
core particle (1kx5), \cite{SI_RichmondNCP} and the corresponding DNABEND predictions (Supplementary Fig. 10).
This procedure yields $w=0.1$ for the 147 bp superhelix and $w=0.5$ for the 71 bp superhelix bound by the
$\mathrm{H3_{2}H4_{2}}$ tetramer (bps 39 through 109 in the 147 bp superhelix).
DNABEND is not very sensitive to the exact value of $w$: we found a correlation of 0.99 between the
free energies computed using $w=0.1$ and $w=0.5$ and the 71 bp superhelix (data not shown).
Note that the total energy is a sum of two quadratic potentials written out in terms of the deviations
of the global displacements $\delta \vec{d}_s$ and the local frame angles
$\delta \vec{\Omega}_s$ from their ideal superhelical values.
Given the value of the fitting weight, the conformation adopted by the DNA molecule is the one that minimizes its total energy $E$.
Since the energy is quadratic, finding the energy minimum is equivalent to solving a system of linear equations:
\begin{equation} \label{Ederiv}
\begin{array}{ccc}
{\partial E}/{\partial \delta \alpha^{s}_{i}} = 0 & s = 1 \dots N, & i = 1 \dots 6.
\end{array}
\end{equation}
The numerical solution of the system of equations (\ref{Ederiv}) provides a set of geometric parameters
corresponding to the minimum DNA energy: $(\delta \vec{\Omega}_s, \delta \vec{d}_s)$. These can be used
to find the original geometric parameters in the local frame:
$(\Omega_s^0 + \delta \vec{\Omega}_s, \vec{d}_s^{0} + {R_s^{MST}}^{-1} \delta \vec{d}_s)$.

\vspace{0.2in}
\noindent \textbf{Worm-like chain.}
According to the worm-like chain model, the energy required to bend a DNA molecule is sequence-independent
and given by: \cite{SI_Bouchiat}
\begin{equation} \label{wlc}
E_{wlc} = \frac{k_B T L_p}{2} \int_{0}^{L_0} ds |\frac{d \vec{\mathbf t}}{ds}|^2,
\end{equation}
where $L_0$ is the contour length of the molecule, $L_p$ is the persistence length (estimated to be $\simeq 400$ \AA), \cite{SI_Bouchiat}
$k_B T \simeq 0.6$ kcal/mol, and $\vec{\mathbf t}$ is the unit tangent vector.
The contour length of the ideal 147 bp superhelix is given by $146 \times 3.333$ \AA.
From Equations (\ref{ideal:s}) and (\ref{tnb}) we obtain
$|d \vec{\mathbf t} / ds|^2 = R^2 / R_{eff}^4$, and thus
\begin{equation*}
E_{wlc} = \frac{k_B T L_p L_0}{2} \frac{R^2}{R_{eff}^4} \simeq 32.6 ~\mathrm{kcal/mol}.
\end{equation*}

In Fig. 2b, the mean and the standard deviation $\sigma$ of DNABEND energies computed for chromosome III are
$127.0 \pm 6.1$. Equating the worm-like chain model estimate with the mean DNABEND energy,
we obtain a scaling coefficient of 0.26 which can be used to express DNABEND energies in kcal/mol. This gives the difference
of 15.2 kcal/mol between the best and the worst chromosome III sequences, and $\sigma = 1.6$ kcal/mol.
Most sequences differ by $2 \sigma$ or less in Fig. 2b and are thus separated by $\le 6.4$ kcal/mol.
A similar value of the scaling coefficient (0.21) arises from a linear model fit between
experimental and DNABEND-predicted free energies in Supplementary Fig. 12a (red circles).

\subsection{Predicting genome-wide occupancies of DNA-binding proteins} \label{DynaPro}


Chromatin structure plays an important role in regulating eukaryotic gene expression. \cite{SI_Khorasanizadeh,SI_Boeger,SI_Wallrath,SI_Jenuwein}
Eukaryotic DNA is packaged by histones into chains of nucleosomes that are subsequently folded into $\sim 30$ nm chromatin fibers.
Because arrays of nucleosomes form on chromosomal DNA under physiological conditions, in order to predict genomic nucleosomal
occupancies we need to describe DNA packaged into multiple non-overlapping nucleosomes. Furthermore,
nucleosomes may compete with other DNA binding proteins for genomic sequence.
For example, several closely spaced TF binding sites
may serve to displace the nucleosomes from the promoter region upon TF binding. To take the competition between
nucleosomes and other factors into account we need to consider configurations with multiple regularly spaced
nucleosomes (accomodating a 20-25 bp linker between the end of one nucleosome and the beginning of the next)
as well as other DNA-binding proteins.
The probability of any such configuration can be computed
if we can construct a statistical sum (partition function) over all possible configurations. Though the sum has exponentially
many terms and is thus impossible to evaluate by brute force for all but very short DNA sequences,
it can nonetheless be efficiently computed with a dynamic programming algorithm. \cite{SI_Durbin}

Here we develop the dynamic programming approach for a general case of
$M$ objects of length $L_j~(j = 1 \dots M)$ that are placed on genomic DNA (\textit{i.e.} occupy $L_j$ DNA
basepairs). The objects could represent nucleosomes, TFs, or any other DNA-binding proteins. The binding energy of object $j$
with DNA at each allowed position $i$ is assumed to be known: $E^{j}_{i}~(i = 1 \dots N-L_{j}+1)$, where $N$ is the number of basepairs.
For nucleosomes we will carry out a genome-wide computation of DNA elastic energies as described above, while for other
factors the energy landscapes will be constructed based on their binding preferences inferred from footprinting experiments,
SELEX assays, bioinformatics predictions, etc.
We assign index $0$ to the background which can be formally considered to be an object of length 1: $L_0 = 1$. 
In the simplest case which we consider here the background energy is zero everywhere, but more sophisticated models could 
incorporate a global bias by making positions near DNA ends less favorable, etc.

We wish to compute a statistical sum over all possible configurations in which object overlap is not allowed (including
the background ``object''):
\begin{equation} \label{Z:part}
Z = \sum_{conf} e^{-E(conf)},
\end{equation}
where $E(conf) = \sum_{j=0}^{M} \sum_{i=1}^{N^{j}_{obj}} E^{j}_{c(i)} $ is the total energy of an arbitrary configuration
of non-overlapping objects, $N^{j}_{obj}$ is the number of objects of type $j$, and $E^{j}_{c(i)}$ is the pre-computed
energy of the object of type $j$ which occupies positions $c(i)$ through $c(i) + L_j - 1$.

It is possible to evaluate $Z$ (or the free energy $F = \log(Z)$) efficiently
by recursively computing the partial statistical sums:
\begin{equation} \label{Z:Forward}
Z^f_i = \sum_{j=0}^{M} Z^f_{i-L_{j}} e^{-E^j_{i-(L_{j}-1)}} \theta_{i-(L_{j}-1)}, ~~i = 1 \dots N,
\end{equation}
with the initial condition $Z^f_0 = 1$. The theta function is defined as:
\begin{equation} \label{theta:def}
\theta_j = \left\{ \begin{array}{l}
1, ~j > 0 \\
0, ~j \le 0
\end{array} \right.
\end{equation}
It is computationally more efficient to transform Equation (\ref{Z:Forward}) into log space by defining the partial
free energies $F_i = \log Z^f_i$:
\begin{equation} \label{F:Forward}
F_i = \log \left( \sum_{j=0}^{M} e^{ F_{i-L_{j}} - E^j_{i-L_{j}+1} } \theta_{i-L_{j}+1} \right), ~~i = 1 \dots N,
\end{equation}
with the initial condition $F_0 = 0$.
For numerical stability, we rewrite Equation (\ref{F:Forward}) as an explicit update of the partial free energy
computed at the previous step:
\begin{equation} \label{F:Forward:final}
F_i = F_{i-1} + \log \left( \sum_{j=0}^{M} e^{ F_{i-L_{j}} - F_{i-1} - E^j_{i-L_{j}+1} } \theta_{i-L_{j}+1} \right), ~~i = 1 \dots N.
\end{equation}
The free energy computed in the final step is the full free energy where all possible configurations are taken into account:
$F_N = F = \log(Z)$. Since the algorithm proceeds by computing partial sums from 1 to N it is often called the forward pass.
Similar equations can be constructed for the backward pass which proceeds from N to 1:
\begin{equation} \label{Z:Backward}
Z^r_i = \sum_{j=0}^{M} Z^r_{i+L_{j}} e^{-E^j_{i}} \theta_{N-i-L_{j}+2}, ~~i = N \dots 1,
\end{equation}
with the initial condition $Z^r_{N+1} = 1$. We transform Equation (\ref{Z:Backward}) into log space by defining
the backward partial free energies $R_i = \log Z^r_i$:
\begin{equation} \label{R:Backward}
R_i = \log \left( \sum_{j=0}^{M} e^{ R_{i+L_{j}} - E^j_{i} } \theta_{N-i-L_{j}+2} \right), ~~i = N \dots 1,
\end{equation}
with the initial condition $R_{N+1} = 0$, or equivalently:
\begin{equation} \label{R:Backward:final}
R_i = R_{i+1} + \log \left( \sum_{j=0}^{M} e^{ R_{i+L_{j}} - R_{i+1} - E^j_{i} } \theta_{N-i-L_{j}+2} \right), ~~i = N \dots 1.
\end{equation}
Note that $R_1 = F_N = F$ by construction.

With the full set of forward and backward partial free energies we can evaluate any statistical quantity of interest.
For example, the probability of finding an object of type $j$ at positions $(i \dots i + L_j - 1)$ is given by:
\begin{equation} \label{Prob:obj}
P^j_i = \frac{Z^f_{i-1} e^{-E^j_i} Z^r_{i+L_j}}{Z} = e^{F_{i-1} - E^j_i + R_{i+L_j} - F}, ~~i = 1 \dots N-L_j+1.
\end{equation}

Another quantity of interest is the occupancy of the basepair $i$ by object $j$, defined as the probability that
basepair $i$ is covered by any object of type $j$: \cite{SI_Segal}
\begin{equation} \label{Occ:def}
O^j_i = \sum_{k = i - (L_j - 1)}^i P^j_k, ~~i = 1 \dots N.
\end{equation}
(note that $P^j_k = 0$ for $k > N-L_j+1$). If the object is composite (\textit{i.e.} consists of $L_j$ basepairs
extended symmetrically on both sides by $L_j^{em}$ to take the nonzero linker lengths into account: $\bar{L}_j = L_j + 2L_j^{em}$),
we may be interested in the partial occupancy by $L_j$ basepairs at the center of the object.
Equation (\ref{Occ:def}) then becomes:
\begin{equation} \label{Occ:embed}
O^j_i = \sum_{k = i - (L_j - 1) - L_j^{em}}^{i-L_j^{em}} P^j_k = \sum_{k = i - (\bar{L}_j - L_j^{em})}^{i-L_j^{em}} P^j_k.
\end{equation}
Recursively,
\begin{equation} \label{Occ:recurs}
O^j_i = O^j_{i-1} + P^j_{i-L_j^{em}} - P^j_{i - (\bar{L}_j - L_j^{em}) - 1}.
\end{equation}
Here we assume that all quantities on the right-hand side are set to zero if their indices are outside
their definition domains (such as $O^j_0$).

Finally, we need to take into account the fact that the objects can bind DNA in both directions, and thus there are
two binding energies for each position: $E^j_i$ (object $j$ starts at $i$ and extends in the 5' to 3' direction)
and $E^{j (rc)}_i$ (object $j$ starts at $i + L_j - 1$ and extends in the 3' to 5' direction). It is easy to show
that the formalism developed above applies without change if the binding energies $E^j_i$ are replaced by the free energies
$\bar{E}^j_i$ which take both binding orientations into account:
\begin{equation} \label{Free_en:both}
\bar{E}^j_i = - \log [ e^{-E^j_i} + e^{-E^{j (rc)}_i} ].
\end{equation}

In the case of a single type of DNA-binding object (such as the nucleosome energies predicted with DNABEND) there are two
free parameters: the mean energy of nucleosomes $\langle E^{nuc} \rangle$ over a chromosome or a given genomic 
region which affects overall nucleosome occupancy, and the standard deviation $\sigma (E^{nuc})$
which plays the role of inverse temperature. Note that DNABEND nucleosomes are 157 bp long because
a 147 bp histone binding site is flanked on both sides by 5 bp long linkers which do not contribute to the energy. 
Because the background energy is set to zero, making $\langle E^{nuc} \rangle$ more negative results in fewer bases
being devoid of nucleosomes (this corresponds to increasing free concentration of histone octamers
in the histone-DNA solution), and vice versa. The inverse temperature is related to the nucleosome stability:
making $\sigma (E^{nuc})$ smaller results in fewer stable nucleosomes (\textit{e.g.} those present in half or more of all
the configurations in Equation (\ref{Z:part})), while allowing for a bigger difference between favorable and unfavorable
nucleosomal energies results in more configurations with 'frozen', stable nucleosomes. For all calculations in this
paper we have chosen to set $\langle E^{nuc} \rangle = 0.0$, $\sigma^2 (E^{nuc}) = 45.0$ for every chromosome.
The nucleosome energies with zero mean result in the average nucleosome occupancy of 0.797 over all chromosomes, and 12538 stable,
non-overlapping nucleosomes (with $P > 0.5$) covering 16.3\% of the yeast genome. In addition, this value of $\sigma (E^{nuc})$
is close to the scale of DNABEND elastic energies. The free parameters described here are also present in the alignment model
(albeit in a more implicit manner, see section \ref{aln:methods}), and are required in general for each type of DNA-binding object.

\subsection{TF and TBP binding energies} \label{TF:TBP}

All TF binding energies were computed using position-specific weight matrices (PWMs) from 
Morozov \textit{et al.} \cite{SI_Morozov_II} and MacIsaac \textit{et al.}: \cite{SI_MacIsaac_I}
\begin{equation} \label{TF:score}
E^{TF}(S) = - C \sum_{i=1}^{L_{TF}} \log \frac{w_{i}^{\alpha_i}}{b^{\alpha_i}},
\end{equation}
where $L_{TF}$ is the length of the binding site, $\alpha_i = \{A,C,G,T\}$ is the nucleotide at position $i$,
$w_{i}^{\alpha_i}$ is the frequency of base $\alpha_i$ at position $i$ in the PWM,
and $b^{\alpha_i}$ is the background frequency of base $\alpha_i$: $\{b^A,b^C,b^G,b^T\} = \{0.309, 0.192, 0.191, 0.308\}$
(the same background frequencies as used in the alignment-based nucleosome model, see section \ref{aln:methods}).
$C$ is an energy scale, set to 1 for all TFs.
The distribution of energies is bounded from below by the energy of the consensus sequence $E_{min}$
which is a free parameter of the model since it depends on the TF concentration in solution.
Thus the energy of a TF bound to site $S$ is given by:
$E^{TF}_{new}(S) = E_{offset} + (E^{TF}(S) - E_{min})$, where $E_{offset}$ is typically set to -6.0 (relative to the 0.0
average energy of DNABEND nucleosomes), but can be also varied
through a range of energies (cf. Fig. 4d). Finally, the two energies of binding the same site in both directions are
combined using:
\begin{equation} \label{TF:score:final}
\bar{E}^{TF}(S) = - \log \left[ B_w e^{-E^{TF}_{new}(S)} + (1 - B_w) e^{-E^{TF}_{new}(S^{(rc)})} \right],
\end{equation}
where $B_w = 0.5$ is the strand bias. A set of TF binding energies computed in this way
(together with the nucleosome binding energy profile) is used as input to the
recursive algorithm from Section \ref{DynaPro}. In practice, we restrict a set of sites to which
TFs are allowed to bind by assigning a highly unfavorable energy to all TF positions that are not
listed as predicted sites by MacIsaac \textit{et al.} \cite{SI_MacIsaac_I}

TBP binding energies were obtained in the same way, except that 
the TBP PWM was derived using an alignment of TATA box sites from Basehoar \textit{et al.} \cite{SI_Basehoar}
TBPs were only allowed to bind the sites listed in Basehoar \textit{et al.} as TATA boxes.

\subsection{Nucleosome positions from tiled microarray analysis} \label{Rando:methods}

Yuan \textit{et al.} have developed a hidden Markov model-based approach for inferring nucleosome positions
from high resolution tiled microarray data. \cite{SI_Yuan}
The data was collected for most of \textit{S.cerevisiae} chromosome III,
plus additional regulatory regions. We have used Yuan \textit{et al.} ``HMM calls'' listed in the
Supplementary Data to identify probes occupied by either a well-positioned or a fuzzy nucleosome
(manually added ``hand calls'' were not considered). We then extended nucleosomal regions
from the mid-point of the first occupied probe to the mid-point of the last occupied probe,
plus additional 10 bp on each side (making it consistent with the definition of nucleosome coverage
of TF binding sites employed in Fig. 3 of Yuan \textit{et al.}). \cite{SI_Yuan}
Thus the nucleosomal length is given by $20(N-1)+20$, where $N$ is the number of contiguous probes
occupied by a single nucleosome.
We find that most nucleosomes occupy 6 contiguous probes and thus cover 120 bp, somewhat less than 147 bp expected from
the structural point of view. The total number of nucleosomes placed on chromosome III is 1045;
every basepair covered by either well-positioned or fuzzy nucleosome was assigned an occupancy of 1.0, resulting
in the overall average occupancy of 0.488 (that is, almost half of the genome is covered by the non-overlapping
nucleosomes). The resulting occupancy profile and the corresponding nucleosome positions
were used to compile all Yuan \textit{et al.} results in this work.

\subsection{H2A.Z nucleosome positions from DNA sequencing} \label{Albert:methods}

Albert \textit{et al.} have used high throughput DNA sequencing to identify genomic positions of H2A.Z
nucleosomes in \textit{S.cerevisiae} with a median 4 bp resolution. \cite{SI_Albert} We converted their nucleosome
positioning data into the occupancy profile by using the ``fine-grain'' nucleosome center
coordinates (reported for both strands in Albert \textit{et al.} Supplementary Materials)
to place 147 bp nucleosomes on the genome.
Since many ``fine-grain'' nucleosomes overlap, we end up with
longer nucleosome-occupied regions containing multiple hits and surrounded by regions with zero nucleosome coverage.
To each base pair in these occupied regions we then assign an integer equal to the number of nucleosomes that cover
this particular base pair, thus producing an unnormalized nucleosome occupancy profile. Finally, the occupancy
profile is normalized by rescaling the integer
counts separately in each occupied region so that the highest occupancy is assigned a value of 1.0.
The total number of overlapping H2A.Z nucleosome reads is 34796; with the conventions described above
the average occupancy is 0.118 over all chromosomes. 
The resulting occupancy profile and the corresponding nucleosome positions
were used to produce Albert \textit{et al.} results for Supplementary Figs. 22 and 27b.

\subsection{Alignment-based model of nucleosome energies and occupancies} \label{aln:methods}

Segal \textit{et al.} have recently developed a nucleosome positioning model based on the alignment of 199 sequences
occupied by nucleosomes \textit{in vivo} in the yeast genome. \cite{SI_Segal} The main assumption of the model
is that it is possible to infer nucleosome formation energies using dinucleotide distributions observed in aligned
nucleosomal sequences. To this end, $N_{seq}$ experimentally determined mononucleosome sequences of length $\sim 147$ bp
and their reverse complements were aligned around their centers as described in Segal \textit{et al.} \cite{SI_Segal}
The rules for choosing the center of the alignment should be clear from the following example:
\begin{verbatim}
      x
AAGCGTTAAACGC  seq1
GCGTTTAACGCTT  seq1_rc
   ATGCGACG    seq2
   CGTCGCAT    seq2_rc
\end{verbatim}
Here 'x' labels the center of the alignment, which is chosen in an obvious way for sequences with an odd number of
nucleotides (seq1), and by making the number of nucleotides to the left of the central nucleotide one less than 
the number of nucleotides to the right of the central nucleotide for sequences with an even number of nucleotides (seq2).

A local average over neighboring dinucleotide positions was carried out by shifting the original alignment of $2 N_{seq}$
by 1 bp to the right as well as 1 bp to the left, and adding the shifted sequences to the original alignment.
Thus the final alignment has $6 N_{seq}$ sequences.
Next, the conditional probability of a nucleotide of type $\alpha$ at position $i$ ($S^{\alpha}_i$)
given a nucleotide of type $\beta$ at position $i-1$ ($S^{\beta}_{i-1}$) is computed at every position
within 67 bp of the center of the alignment:
\begin{eqnarray} \label{Prob:cond}
P (S^{\alpha}_i|S^{\beta}_{i-1}) = \frac{N (S^{\alpha}_i S^{\beta}_{i-1})}{ \sum_{\gamma=1}^{4} N (S^{\alpha}_i S^{\gamma}_{i-1})}, & i = -67 \dots 67
\end{eqnarray}
where $\alpha,\beta,\gamma = \{A,C,G,T\}$, $N (S^{\alpha}_i S^{\beta}_{i-1})$ is the number of times the $\beta\alpha$
dinucleotide occurs in the augmented alignment, and $0$ marks the alignment center.
Keeping conditional probabilities only for 135 central positions disregards dinucleotide counts from the
outer edges of the alignment to which fewer sequences contribute.
The final model is 157 bp long because 135 conditional probabilities are flanked on each side by 11 background
nucleotide frequencies, to approximate steric replusion between nucleosomes:
\begin{eqnarray} \label{P:bkgr}
P (S^{\alpha}_j|S^{\beta}_{j-1}) = P^{g} (S^{\alpha}), & i = (-78 .. -68, 68 .. 78)
\end{eqnarray}
where $\{P^{g}(A),P^{g}(C),P^{g}(G),P^{g}(T)\} = \{0.309012, 0.191674, 0.191303, 0.308012\}$ are
the background frequencies from the yeast genome.
Finally, the model assigns a negative log score (interpreted as the energy of nucleosome formation)
to a 157 bp long nucleotide sequence $S$:
\begin{eqnarray} \label{LogScore}
L^{m}(S) = - \sum_{i=-78}^{78} \log \frac{P(S^{\alpha_i}_i|S^{\alpha_{i-1}}_{i-1})}{P^{m}(S^{\alpha_i})},
\end{eqnarray}
where $\alpha_i$ denotes the nucleotide type at position $i$ in the sequence $S$,
and $P^{m}(S^{\alpha})$ are either the genomic background frequencies defined above or uniform background
frequencies: $\{P^{u}(A),P^{u}(C),P^{u}(G),P^{u}(T)\} = \{0.25, 0.25, 0.25, 0.25\}$ (\textit{i.e.} $m = \{g,u\}$).
Note that by construction the flanking nucleotides do not contribute to Equation (\ref{LogScore}) (apart from a
possible difference in background frequencies), and thus play the role of embedding linker regions.
Finally, to reproduce results from Segal \textit{et al.} \cite{SI_Segal}
the energies of both strands are combined using an
empirical formula which implicitly sets the temperature,
resulting in 13978 stable, non-overlapping nucleosomes (with probability $P \ge 0.5$) that cover 18.2\% of the yeast genome,
and the average occupancy of 0.844 over all chromosomes:
\begin{equation} \label{LogScore:combined}
L^{I}(S_k) = L^{g}(S_k) + L^{u}(S_k^{(rc)}),
\end{equation}
where $S_k$ is a 157 bp sequence starting at position $k$ in the genome, and $S_k^{(rc)}$ is its reverse complement.
Note that genomic background frequencies are used for one strand while uniform background frequencies
are used for the other.
We refer to the log scores computed in this way, and the corresponding nucleosome probabilities and occupancies
as alignment model I. Alternatively, in alignment model II we use genomic background frequencies for both
DNA strands and rescale the temperature explicitly to match DNABEND and alignment model I:
\begin{equation} \label{LogScore:rescaled}
L^{II}(S_k) = 1.4 [L^{old}(S_k) - \langle L^{old}(S_k) \rangle] + \langle L^{old}(S_k) \rangle,
\end{equation}
where $L^{old}(S_k) = L^{g}(S_k) + L^{g}(S_k^{(rc)})$
and $\langle \dots \rangle$ signifies an average over the chromosome.
Note that while the log scores are rescaled in Equation (\ref{LogScore:rescaled}),
their mean stays the same (and close to zero because the background scores have been subtracted, cf. Equation (\ref{LogScore})).
We treat the log scores from Equations (\ref{LogScore:combined}) and (\ref{LogScore:rescaled})
as nucleosome energies, and use them to
compute nucleosome probabilities and occupancies as described in section \ref{DynaPro}.
We find that alignment model II predictions do not strongly depend on the exact value of the
scaling coefficient.
Alignment model II results in the average nucleosome occupancy of 0.810 over all chromosomes; 26670 stable,
non-overlapping nucleosomes (with $P \ge 0.5$) are placed, covering 34.7\% of the yeast genome.
The overall correlation between alignment models I and II is 0.66 for log scores and 0.60 for occupancies.

\subsection{Histone-DNA binding affinity measurements} \label{dialysis}

We used a standard competitive nucleosome reconstitution procedure to measure the relative affinity of different
DNA sequences for binding to histones in nucleosomes. \cite{SI_Thastrom_III} In this method, differing tracer DNA molecules compete
with an excess of unlabeled competitor DNA for binding to a limiting pool of histone octamer. 
The competition is established in elevated [NaCl], such that histone-DNA interaction affinities are suppressed
and the system equilibrates freely. The [NaCl] is then slowly reduced by dialysis, allowing nucleosomes to form;
further reduction in [NaCl] to physiological concentrations or below then "freezes-in" the resulting equilibrium,
allowing subsequent analysis, by native gel electrophoresis, of the partitioning of each tracer between free DNA
and nucleosomes. The distribution of a given tracer between free DNA and nucleosomes defines an equilibrium constant
and a corresponding free energy, valid for that competitive environment. Comparison of the results for a given pair
of tracer DNAs (in the identical competitive environment) eliminates the dependence on the details of the competitive
environment, yielding the free energy difference ($\Delta \Delta G$) of histone-interaction between the two tracer DNAs.
To allow for comparison with other work, we include additional tracer DNAs as reference molecules:
a derivative of the 5S rDNA natural nucleosome positioning sequence \cite{SI_Thastrom_III} and the 147 bp nucleosome-wrapped region
of the selected high affinity non-natural DNA sequence 601. \cite{SI_Thastrom_II}

The 5S and 601 reference sequences were prepared by PCR using plasmid clones as template.
146 and 147 bp long DNAs analyzed in X-ray crystallographic studies of nucleosomes (PDB
entries 1aoi and 1kx5, respectively) were prepared as described \cite{SI_Dyer} using clones supplied
by Professors K. Luger and T.J. Richmond, respectively. New 147 bp long DNA sequences
designed in the present study were prepared in a two-step PCR-based procedure using
chemically synthesized oligonucleotide primers. All synthetic oligonucleotides were gel purified
prior to use. The central 71 bp were prepared by annealing the two strands. The resulting duplex
was gel purified and used as template in a second stage PCR reaction to extend the length on
each end creating the final desired 147 bp long DNA. The resulting DNA was again purified by
gel electrophoresis.

DNA sequences to be analyzed were 5' end-labeled with $^{32}$P, and added in tracer quantities to
competitive nucleosome reconstitution reactions. Reconstitution reactions were carried out as
described \cite{SI_Thastrom_III} except that each reaction included 10 $\mu$g purified histone octamer and 30 $\mu$g
unlabeled competitor DNA (from chicken erythrocyte nucleosome core particles) in the 50 $\mu$l
microdialysis button.

\subsection{Hydroxyl radical footprinting of nucleosomal templates} \label{601_603_605}

\textbf{DNA templates.} Plasmids pGEM-3Z/601, pGEM-3Z/603 and pGEM-3Z/605 containing nucleosome positioning sequences
601, 603 and 605, respectively, were described previously. \cite{SI_Lowary}
To obtain templates for hydroxyl radical footprinting
experiments the desired $\sim 200$-bp DNA fragments were PCR-amplified using various pairs of primers and Taq DNA
polymerase (New England BioLabs). The sequences of the primers will be provided on request. To selectively label
either upper or lower DNA strands one of the primers in each PCR reaction was 5'-end radioactively labeled with
polynucleotide kinase and $\gamma^{32}$P-ATP. \cite{SI_Walter}
The single-end-labeled DNA templates were gel-purified and single
nucleosomes were assembled on the templates by dialysis from 2 M NaCl. \cite{SI_Walter}
Nucleosome positioning was unique on at least 95\% of the templates.

\textbf{Hydroxyl radical footprinting.} Hydroxyl radicals introduce non-sequence-specific single-nucleotide gaps in DNA,
unless DNA is protected by DNA-bound proteins. \cite{SI_Tullius}
Hydroxyl radical footprinting was conducted using single-end-labeled
histone-free DNA or nucleosomal templates as previously described. \cite{SI_Tullius}
In short, 20-100 ng of single-end-labeled DNA or nucleosomal
templates were incubated in 10 mM HEPES buffer (pH 8.0) in the presence of hydroxyl radical-generating reagents
present at the following final concentrations (2 mM Fe(II)-EDTA, 0.6\% H2O2, 20 mM Na-ascorbate) for 2 min at
$20^{\circ}$~C. Reaction was stopped by adding thiourea to 10 mM final concentration.
DNA was extracted with Phe:Chl (1:1), precipitated with ethanol, dissolved in a loading buffer and analyzed
by 8\% denaturing PAGE.

\textbf{Data analysis and sequence alignment.} The denaturing gels were dried on Whatman 3MM paper, exposed to
a Cyclone screen, scanned using a Cyclone and quantified using OptiQuant software (Perklin Elmer).
Positions of nucleotides that are sensitive to or protected from hydroxyl radicals were identified by
comparison with the sequence-specific DNA markers (Supplementary Fig. 13). The dyad was localized by
comparison of the obtained footprints with the footprints of the nucleosome assembled on human
$\alpha$-satellite DNA. The latter footprints were modeled based on the available 2.8 \AA~ resolution
X-ray nucleosome structure. \cite{SI_LugerNCP}


\newpage

\section{Supplementary Tables}

\renewcommand{\baselinestretch}{1.1}   

{\normalsize
\noindent
\textbf{Table 1.} Predicted ($\mathbf{F_{pred}}$) and measured ($\mathbf{F_{exp}}$)
free energies of computationally designed sequences.
Experimental free energies are shown relative to the reference sequence
from the \textit{L.variegatus} 5S rRNA gene. Histone binding is dominated by the contribution from
the $\mathrm{H3_{2}H4_{2}}$ tetramer with the 71 bp binding site. \cite{SI_Thastrom_II}
The best binder is created by using simulated annealing to introduce mutations and thus
minimize the energy of a 71 bp DNA molecule.
B71S1 and B71S2 have different sequences flanking the 71 bp designed site
(whose contribution dominates the total free energy).
601S1 and 601S2 consist of the 71 bp site from the center of the 601 sequence \cite{SI_Lowary}
and flanking sequences from B71S1 and B71S2, respectively.
W147S is a 147 bp sequence whose free energy (with contributions from multiple $\mathrm{H3_{2}H4_{2}}$ binding sites)
was maximized by simulated annealing. X146 and X147 are 146 bp and 147 bp DNA sequences from nucleosome crystal
structures 1aoi \cite{SI_LugerNCP} and 1kx5.\cite{SI_RichmondNCP}
}

{\normalsize
\begin{center}
\begin{tabular}{|c|c|c|} \hline
 & $\mathbf{F_{pred}}$ (arb.units) & $\mathbf{F_{exp}}$ (kcal/mol) \\ \hline \hline
B71S1 & -20.09 & -1.57 $\pm$ 0.41 \\ \hline
B71S2 & -20.10 & -1.51 $\pm$ 0.27 \\ \hline
601S1 & -0.49 & -2.99 $\pm$ 0.55 \\ \hline
601S2 & -1.74 & -2.46 $\pm$ 0.18 \\ \hline
W147S & 15.26 &  0.09 $\pm$ 0.23 \\ \hline
X147  & 5.92 &  0.75 $\pm$ 0.29 \\ \hline
X146  & 6.86 &  0.45 $\pm$ 0.91 \\ \hline 
\end{tabular}
\end{center}
}

\newpage
{\normalsize
\noindent
\textbf{Table 2.} Average values of DNA geometric parameters for each dinucleotide type.
Dinucleotides are counted in both 5' and 3' directions of strand I (which is chosen arbitrarily).
Note that by construction only tilt and shift switch signs when the dinucleotide
direction is changed. \cite{SI_SchnapsI,SI_SchnapsII}
N - total number of dinucleotides of a given type found in the non-redundant
database of protein-DNA structures (see Supplementary Methods).
Angles (twist, roll, tilt) are measured in degrees ($^\circ$), displacements (shift, slide, rise) are measured
in Angstroms (\AA).
}

{\normalsize
\begin{center}
\begin{tabular}{|c|c|c|c|c|c|c|c|} \hline
Dinucleotide & N & Twist & Roll & Tilt & Shift & Slide & Rise \\ \hline \hline
AA/TT &  163 &  35.31 &  0.76 & $\mp$1.84 & $\mp$0.05 & -0.21 &  3.27 \\ \hline
AC/GT &  154 &  31.52 &  0.91 & $\mp$0.64 & $\pm$0.21 & -0.54 &  3.39 \\ \hline
AG/CT &  128 &  33.05 &  3.15 & $\mp$1.48 &  $\pm$0.12 & -0.27 &  3.38 \\ \hline
CA/TG &  146 &  35.02 &  5.95 & $\mp$0.05 & $\mp$0.16 &  0.18 &  3.38 \\ \hline
CC/GG &  133 &  33.17 &  3.86 &  $\pm$0.40 &  $\pm$0.02 & -0.47 &  3.28 \\ \hline
GA/TC &  138 &  34.80 &  3.87 & $\mp$1.52 & $\mp$0.27 & -0.03 &  3.35 \\ \hline
CG &  124 &  35.30 &  4.29 &  0.00 &  0.00 &  0.57 &  3.49 \\ \hline
GC &  102 &  34.38 &  0.67 &  0.00 &  0.00 & -0.07 &  3.38 \\ \hline
AT &  174 &  31.21 & -1.39 &  0.00 &  0.00 & -0.56 &  3.39 \\ \hline
TA &  156 &  36.20 &  5.25 &  0.00 &  0.00 &  0.03 &  3.34 \\ \hline
\end{tabular}
\end{center}
}

{\normalsize
\noindent
\textbf{Table 3.} Standard deviations of DNA geometric parameters for each dinucleotide type.
Dinucleotides are counted in both 5' and 3' directions of strand I (which is chosen arbitrarily).
N - total number of dinucleotides of a given type found in the non-redundant
database of protein-DNA structures (see Supplementary Methods).
}

{\normalsize
\begin{center}
\begin{tabular}{|c|c|c|c|c|c|c|c|} \hline
Dinucleotide & N & Twist & Roll & Tilt & Shift & Slide & Rise \\ \hline \hline
AA/TT  & 163/163 & 3.61 &  4.55 &  1.69 &  0.32 &  0.39 &  0.19 \\ \hline
AC/GT & 154/154 & 3.10 &  3.91 &  1.85 &  0.48 &  0.32 &  0.20 \\ \hline
AG/CT & 128/128 & 4.06 &  4.07 &  2.27 &  0.51 &  0.51 &  0.24 \\ \hline
CA/TG & 146/146 & 4.17 &  4.67 &  2.15 &  0.56 &  0.77 &  0.26 \\ \hline
CC/GG & 133/133 & 4.76 &  4.37 &  2.86 &  0.67 &  0.73 &  0.27 \\ \hline
GA/TC & 138/138 & 4.03 &  3.78 &  2.06 &  0.48 &  0.61 &  0.20 \\ \hline
CG & 124 & 4.49 &  5.54 &  2.42 &  0.69 &  0.56 &  0.28 \\ \hline
GC & 102 & 4.94 &  4.37 &  2.42 &  0.65 &  0.60 &  0.20 \\ \hline
AT & 174 & 3.04 &  3.64 &  1.25 &  0.39 &  0.27 &  0.18 \\ \hline
TA & 156 & 7.07 &  6.17 &  2.35 &  0.48 &  0.91 &  0.21 \\ \hline
\end{tabular}
\end{center}
}

\newpage
{\normalsize
\noindent
\textbf{Table 4.} Force constant matrices for each dinucleotide type. Dinucleotides are counted
in both 5' and 3' directions of strand I (which is chosen arbitrarily). By construction,
tilt and shift switch signs when the dinucleotide direction is changed, \cite{SI_SchnapsI,SI_SchnapsII}
inducing a corresponding sign change in the force constants involving these degrees of freedom.
}

{\normalsize
\begin{center}
\begin{tabular}{|c|c|c|c|c|c|c|} \hline
& Twist & Roll & Tilt & Shift & Slide & Rise \\ \hline \hline
\multicolumn{7}{|c|}{AA/TT} \\ \hline
Twist & 0.130 & 0.041 & $\pm$0.046 & $\pm$0.159 & -0.305 & 0.739 \\ \hline
 Roll & 0.041 & 0.069 & $\pm$0.026 & $\pm$0.053 & -0.213 & -0.135 \\ \hline
 Tilt & $\pm$0.046 & $\pm$0.026 & 0.406 & -0.408 & $\mp$0.073 & $\mp$0.403 \\ \hline
Shift & $\pm$0.159 & $\pm$0.053 & -0.408 & 11.948 & $\pm$2.162 & $\mp$2.511 \\ \hline
Slide & -0.305 & -0.213 & $\mp$0.073 &  $\pm$2.162 & 8.241 & -5.353 \\ \hline
 Rise & 0.739 & -0.135 & $\mp$0.403 & $\mp$2.511 & -5.353 & 35.715 \\ \hline \hline
\multicolumn{7}{|c|}{AC/GT} \\ \hline
Twist & 0.179 & 0.053 & $\pm$0.015 & $\mp$0.083 & -0.118 & 1.244 \\ \hline
 Roll & 0.053 & 0.085 & $\pm$0.003 & $\pm$0.048 & 0.049 & 0.089 \\ \hline
 Tilt & $\pm$0.015 & $\pm$0.003 & 0.323 & -0.312 & $\mp$0.298 & $\mp$0.219 \\ \hline
Shift & $\mp$0.083 & $\pm$0.048 & -0.312 & 4.912 & $\pm$1.672 & $\mp$0.872 \\ \hline
Slide & -0.118 & 0.049 & $\mp$0.298 & $\pm$1.672 & 10.089 & -3.277 \\ \hline
 Rise & 1.244 & 0.089 & $\mp$0.219 & $\mp$0.872 & -3.277 & 35.709 \\ \hline \hline
\multicolumn{7}{|c|}{AG/CT} \\ \hline
Twist & 0.113 & 0.038 & $\mp$0.001 & $\pm$0.168 & 0.040 & 0.850 \\ \hline
 Roll & 0.038 & 0.077 & $\pm$0.022 & $\pm$0.006 & -0.023 & 0.067 \\ \hline
 Tilt & $\mp$0.001 & $\pm$0.022 & 0.280 & -0.365 & $\pm$0.252 & $\mp$0.995 \\ \hline
Shift & $\pm$0.168 & $\pm$0.006 & -0.365 & 4.954 & $\pm$0.527 & $\pm$0.093 \\ \hline
Slide & 0.040 & -0.023 & $\pm$0.252 & $\pm$0.527 & 4.516 & -2.966 \\ \hline
 Rise & 0.850 & 0.067 & $\mp$0.995 & $\pm$0.093 & -2.966 & 29.330 \\ \hline \hline
\multicolumn{7}{|c|}{CA/TG} \\ \hline
Twist & 0.098 & 0.027 & $\mp$0.034 & $\pm$0.061 & -0.254 & 0.799 \\ \hline
 Roll & 0.027 & 0.059 & $\mp$0.046 & $\pm$0.165 & -0.016 & 0.202 \\ \hline
 Tilt & $\mp$0.034 & $\mp$0.046 & 0.393 & -0.965 & $\pm$0.174 & $\mp$0.593 \\ \hline
Shift & $\pm$0.061 & $\pm$0.165 & -0.965 & 5.740 & $\mp$0.117 & $\pm$1.963 \\ \hline
Slide & -0.254 & -0.016 & $\pm$0.174 & $\mp$0.117 & 2.772 & -4.449 \\ \hline
 Rise & 0.799 & 0.202 & $\mp$0.593 & $\pm$1.963 & -4.449 & 23.870 \\ \hline
\end{tabular}
\end{center}
}

\newpage
{\normalsize
\noindent
\textbf{Table 4 (continued).}
}

{\normalsize
\begin{center}
\begin{tabular}{|c|c|c|c|c|c|c|} \hline
& Twist & Roll & Tilt & Shift & Slide & Rise \\ \hline \hline
\multicolumn{7}{|c|}{CC/GG} \\ \hline
Twist & 0.114 & 0.045 & $\pm$0.008 & $\mp$0.139 & -0.328 & 1.587 \\ \hline
 Roll & 0.045 & 0.075 & $\mp$0.014 & $\pm$0.005 & -0.083 & 0.797 \\ \hline
 Tilt & $\pm$0.008 & $\mp$0.014 & 0.218 & -0.537 & $\mp$0.215 & $\pm$0.066 \\ \hline
Shift & $\mp$0.139 & $\pm$0.005 & -0.537 & 3.917 & $\pm$0.085 & $\mp$0.401 \\ \hline
Slide & -0.328 & -0.083 & $\mp$0.215 & $\pm$0.085 & 3.795 & -7.972 \\ \hline
 Rise & 1.587 & 0.797 & $\pm$0.066 & $\mp$0.401 & -7.972 & 41.678 \\ \hline \hline
\multicolumn{7}{|c|}{GA/TC} \\ \hline
Twist & 0.133 & 0.055 & $\pm$0.027 & $\pm$0.073 & -0.350 & 1.301 \\ \hline
 Roll & 0.055 & 0.097 & $\pm$0.042 & $\mp$0.099 & -0.158 & 0.336 \\ \hline
 Tilt & $\pm$0.027 & $\pm$0.042 & 0.408 & -1.012 & $\pm$0.016 & $\pm$0.139 \\ \hline
Shift & $\pm$0.073 & $\mp$0.099 & -1.012 & 10.434 & $\pm$3.814 & $\mp$1.105 \\ \hline
Slide & -0.350 & -0.158 & $\pm$0.016 & $\pm$3.814 & 6.278 & -7.676 \\ \hline
 Rise & 1.301 & 0.336 & $\pm$0.139 & $\mp$1.105 & -7.676 & 41.988 \\ \hline \hline
\multicolumn{7}{|c|}{CG} \\ \hline
Twist & 0.101 & 0.020 & 0.000 & 0.000 & -0.278 & 0.806 \\ \hline
 Roll & 0.020 & 0.040 & 0.000 & 0.000 & 0.001 & -0.046 \\ \hline
 Tilt & 0.000 & 0.000 & 0.255 & -0.510 & 0.000 & 0.000 \\ \hline
Shift & 0.000 & 0.000 & -0.510 & 3.104 & 0.000 & 0.000 \\ \hline
Slide & -0.278 & 0.001 & 0.000 & 0.000 & 3.991 & -2.936 \\ \hline
 Rise & 0.806 & -0.046 & 0.000 & 0.000 & -2.936 & 20.174 \\ \hline \hline
\multicolumn{7}{|c|}{GC} \\ \hline
Twist & 0.069 & 0.006 & 0.000 & 0.000 & -0.217 & 0.646 \\ \hline
 Roll & 0.006 & 0.057 & 0.000 & 0.000 & 0.094 & 0.168 \\ \hline
 Tilt & 0.000 & 0.000 & 0.256 & -0.542 & 0.000 & 0.000 \\ \hline
Shift & 0.000 & 0.000 & -0.542 & 3.473 & 0.000 & 0.000 \\ \hline
Slide & -0.217 & 0.094 & 0.000 & 0.000 & 4.030 & 1.322 \\ \hline
 Rise & 0.646 & 0.168 & 0.000 & 0.000 & 1.322 & 34.392 \\ \hline \hline
\multicolumn{7}{|c|}{AT} \\ \hline
Twist & 0.189 & 0.068 & 0.000 & 0.000 & 0.111 & 1.195 \\ \hline
 Roll & 0.068 & 0.124 & 0.000 & 0.000 & 0.438 & -0.397 \\ \hline
 Tilt & 0.000 & 0.000 & 0.641 & -0.043 & 0.000 & 0.000 \\ \hline
Shift & 0.000 & 0.000 & -0.043 & 6.670 & 0.000 & 0.000 \\ \hline
Slide & 0.111 & 0.438 & 0.000 & 0.000 & 15.942 & -2.611 \\ \hline
 Rise & 1.195 & -0.397 & 0.000 & 0.000 & -2.611 & 47.789 \\ \hline
\end{tabular}
\end{center}
}

\newpage
{\normalsize
\noindent
\textbf{Table 4 (continued).}
}
{\normalsize
\begin{center}
\begin{tabular}{|c|c|c|c|c|c|c|} \hline
& Twist & Roll & Tilt & Shift & Slide & Rise \\ \hline \hline
\multicolumn{7}{|c|}{TA} \\ \hline
Twist &  0.112 &  0.056 &  0.000 &  0.000 & -0.572 &  1.466 \\ \hline
 Roll &  0.056 &  0.064 &  0.000 &  0.000 & -0.146 &  0.318 \\ \hline
 Tilt &  0.000 &  0.000 &  0.365 & -1.271 &  0.000 &  0.000 \\ \hline
Shift &  0.000 &  0.000 & -1.271 &  8.767 &  0.000 &  0.000 \\ \hline
Slide & -0.572 & -0.146 &  0.000 &  0.000 & 4.961 & -12.092 \\ \hline
 Rise &  1.466 &  0.318 &  0.000 &  0.000 & -12.092 & 54.957 \\ \hline
\end{tabular}
\end{center}
}

\newpage

\renewcommand{\figurename}{Supplementary Figure}

\begin{figure}[tbph]
\centering
\includegraphics[scale=0.6]{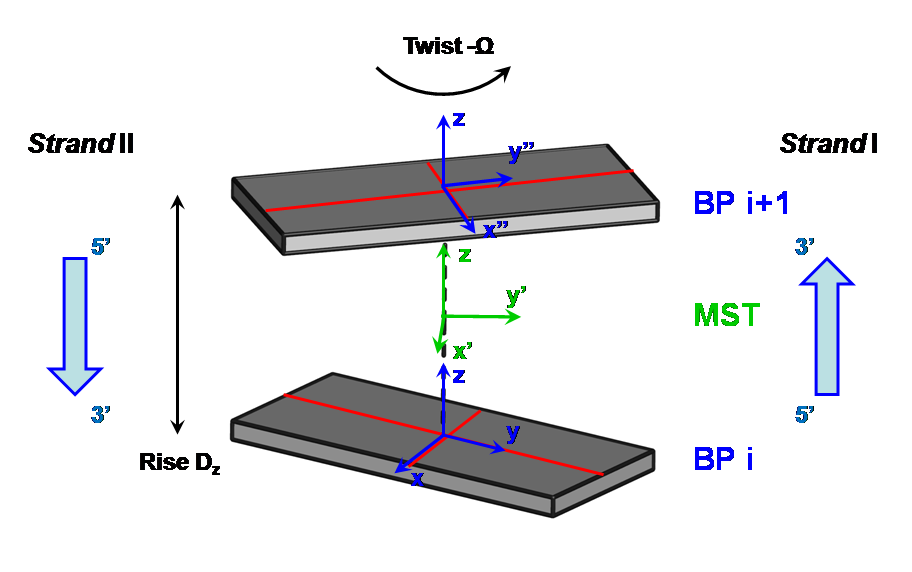}
\caption{
\textbf{Schematic illustration of a single dinucleotide (base step) geometry.}
Coordinate frames attached to basepairs $i$ and $i+1$ are shown in blue,
and the MST coordinate frame is shown in green. For illustrative purposes, only rise $D_z$ and twist $\Omega$ are set
to non-zero values. The origin of the MST frame is at the midpoint of the line connecting the origins of two base pair frames
(which are separated by $D_z$~\AA~ along the z-axis); the MST frame is rotated through $\Omega/2$ with respect to the frame $i$.
}
\end{figure}

\begin{figure}[tbph]
\centering
\includegraphics[scale=0.70]{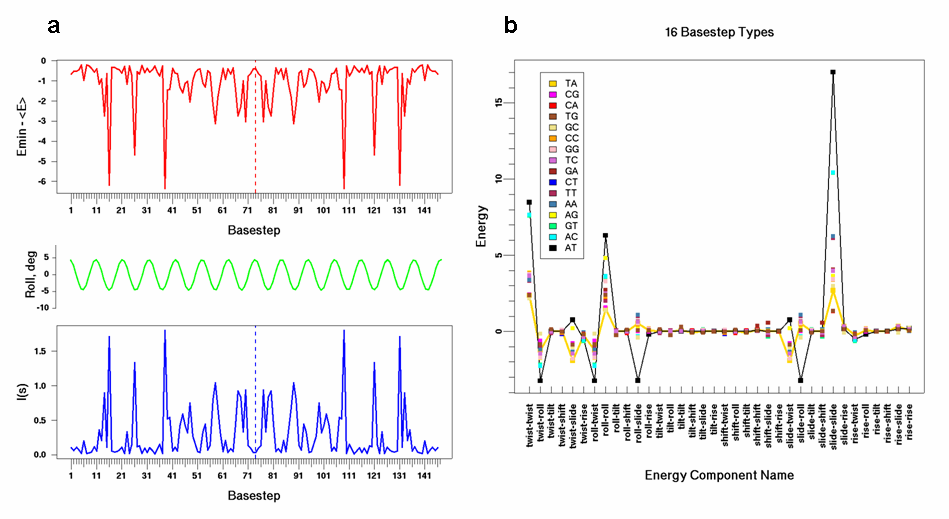}
\end{figure}

\begin{figure}[tbph]
\centering
\caption{
\textbf{Elastic energy analysis based on the DNA conformation from the nucleosome crystal structure.}
a) Position-dependent sequence specificity in the nucleosomal DNA revealed by the energetic
analysis of dinucleotides substituted into the crystal structure of the nucleosome core particle
(PDB code: 1kx5). \cite{SI_RichmondNCP}
All possible dinucleotides were introduced at every position in the 147 bp nucleosomal site
using DNA dihedral angles from the native dinucleotide, and DNA elastic energy was computed for every sequence variant.
Upper panel: the difference between the energy of the most favorable dinucleotide
and the average energy of all dinucleotides at this position.
Lower panel: information score, defined as
$I(s) = \log_2(16) + \sum_{i=1}^{16} p_i^s \log_2(p_i^s)$, where $p_i^s = \exp(-E_i^s)/\sum_{i=1}^{16} \exp(-E_i^s)$,
and $E_i^s$ is the elastic energy change which results from introducing dinucleotide of type $i = 1 .. 16$
at position $s$: $E_i^s = E_i^{s (mut)} - E_i^{s (wt)}$. To enforce the two-fold symmetry of the nucleosome core particle,
all dinucleotide energies were symmetrized around the middle of the DNA site (shown as a dashed vertical line). 
Middle panel: roll angle of the ideal superhelix showing DNA geometry in relation to the histone octamer -
by construction, negative roll angles correspond to the minor groove facing the histones.
b) Elastic energy components for all possible dinucleotides substituted into
the 1kx5 crystal structure at position 109 where the DNA conformation is kinked (Supplementary Fig. 10). \cite{SI_RichmondNCP}
Dinucleotides are ranked by their total energy as shown in the legend (lowest to highest energy from top to bottom).
TA is the lowest energy dinucleotide (thick golden line) and is more favorable than the CA
dinucleotide from the native structure. The energy component analysis reveals that it is the degrees of freedom related to slide
(slide-slide and slide-twist components) and roll (roll-roll component)
that make the TA dinucleotide most favorable, although the slide-slide
component is slightly better in the native CA/TG dinucleotide (red/brown dots).
In contrast, the AT dinucleotide has the highest energy due to its low flexibility with respect to roll, slide, and twist.
The dinucleotide ranking is in agreement with the averages, standard deviations, and force constants
of the roll, slide and twist degrees of freedom observed in the structural database of protein-DNA complexes
(Supplementary Tables 2-4).
}
\end{figure}

\begin{figure}[tbph]
\centering
\includegraphics[scale=0.70]{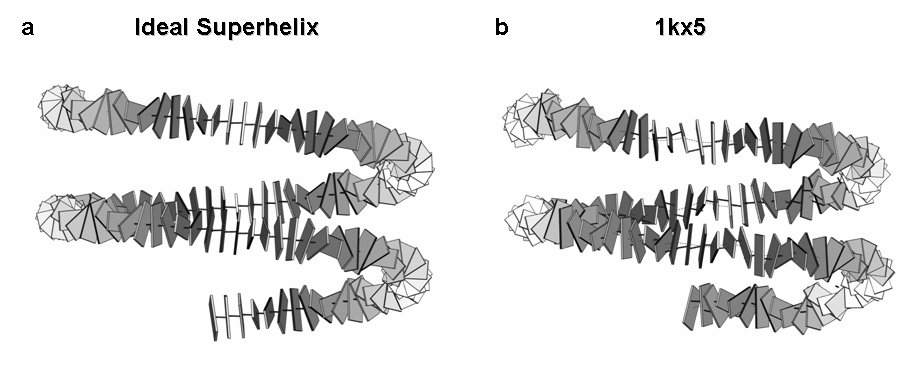}
\caption{
\textbf{Schematic representation of the nucleosomal DNA conformation.} (a) - ideal superhelix,
(b) - high-resolution crystal structure of the nucleosome core particle (PDB code: 1kx5). \cite{SI_RichmondNCP}
Each DNA base pair is shown as a rectangular block.
Images of DNA conformations were created using 3DNA software. \cite{SI_Lu_3DNA}
}
\end{figure}

\begin{figure}[tbph]
\centering
\includegraphics[scale=0.72]{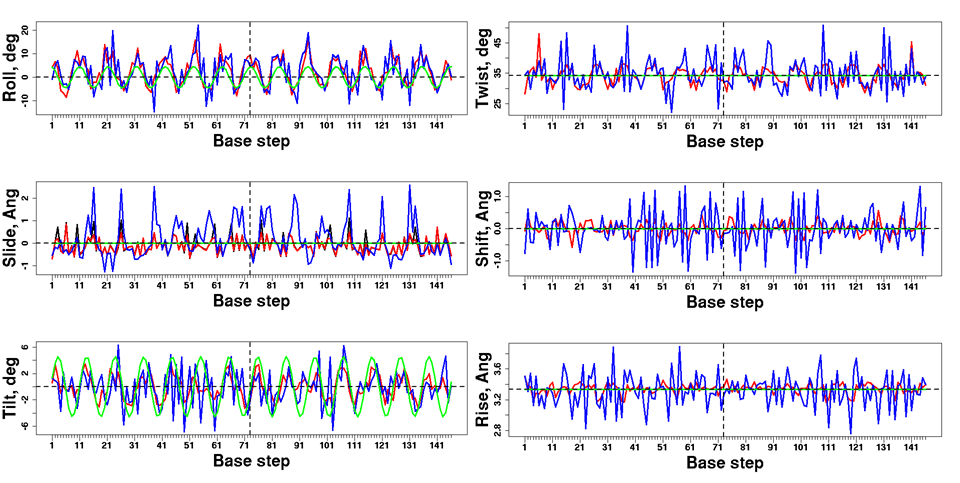}
\caption{
\textbf{DNABEND-predicted and experimentally observed DNA geometries are correlated.}
Distributions of 6 DNA geometric degrees of freedom in the crystal structure of the nucleosome core particle \cite{SI_RichmondNCP}
(PDB code: 1kx5; Supplementary Fig. 9b) (blue), in the minimum energy structure obtained starting from the ideal superhelix
and using 1kx5 DNA sequence as input to DNABEND (red), and in the ideal superhelix with no energy
relaxation (Supplementary Fig. 9a) (green).
The two-fold nucleosome symmetry axis is shown using dashed vertical lines. Mean values of
geometric degrees of freedom in the ideal superhelix are shown as dashed horizontal lines.
Correlation coefficients between the geometric distributions from the native and minimized structures are:
$(r_{twist},r_{roll},r_{tilt},r_{slide},r_{shift},r_{rise}) = (0.489, 0.709, 0.539, 0.536, 0.247, 0.238)$
($\langle r \rangle = 0.460$).
Black curve in the slide panel is a DNABEND prediction based on a modified elastic potential in which the mean
slide for the CA dinucleotide was increased from 0.18
(inferred from a set of non-nucleosome protein-DNA complexes and used in this work)
to 0.91 (inferred from a limited set of available nucleosome structures).
Note that the CA dinucleotides occur at positions (3,10,16,26,50,53,60,66,77,85,102,109,115,133) in the 1kx5
crystal structure. \cite{SI_RichmondNCP}
Whereas the slide correlation coefficient
between the native and minimized structures remained essentially the same at 0.533, the absolute magnitude of predicted
slide peaks is in better correspondence with the observed values. Thus an elastic potential trained on non-nucleosome
complexes may not reproduce all structural aspects of highly bent nucleosomal DNA.
}
\end{figure}

\begin{figure}[tbph]
\centering
\includegraphics[scale=0.75]{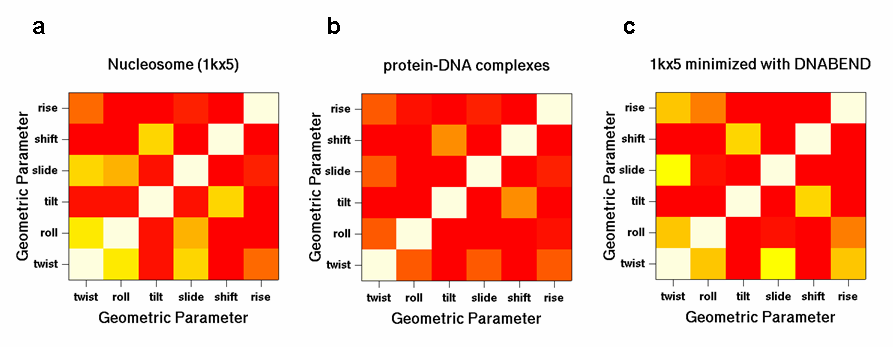}
\caption{
\textbf{DNABEND reproduces experimentally observed couplings between DNA degrees of freedom.}
DNA geometric degrees of freedom are strongly coupled in nucleosomal DNA compared with other protein-DNA complexes.
Shown as heat maps are the matrices of absolute values of correlation coefficients for DNA geometric degrees of freedom
(white: correlation of 1.0, red: little or no correlation).
a) Correlations between DNA geometric parameters observed in the crystal structure
of the nucleosome core particle (PDB code: 1kx5). \cite{SI_RichmondNCP}
b) Correlations found in the structural database of non-homologous protein-DNA complexes \cite{SI_Morozov}
used to build the elastic energy potential.
c) Correlations found in the DNA conformation predicted by DNABEND starting from the ideal superhelix and
using 1kx5 DNA sequence as input. Note that many off-diagonal couplings observed in (a) are reproduced.
}
\end{figure}

\clearpage

\begin{figure}[tbph]
\centering
\includegraphics[scale=0.80]{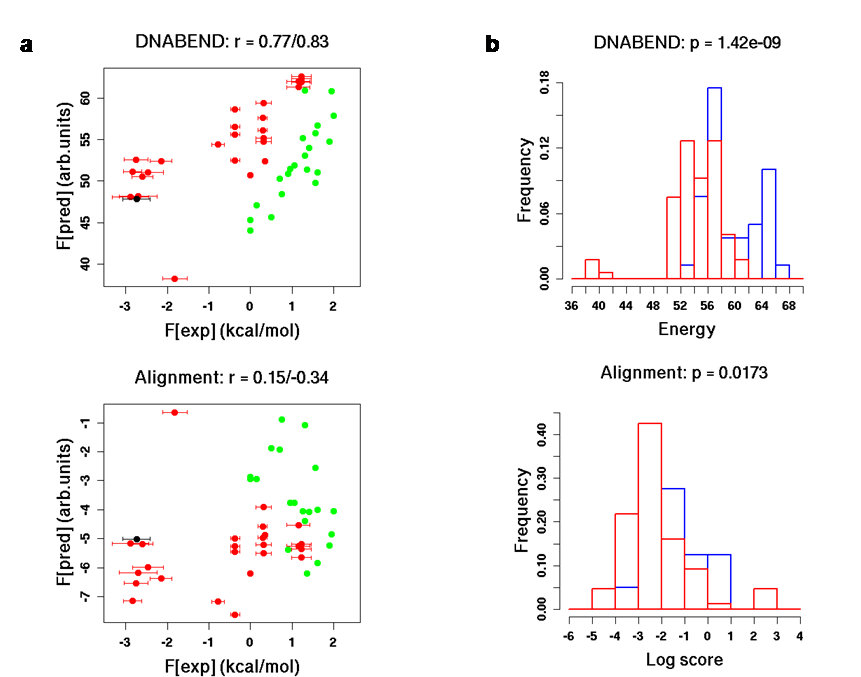}
\end{figure}

\begin{figure}[tbph]
\centering
\caption{
\textbf{DNABEND accurately predicts free energies of nucleosome formation and ranks nucleosome sequences.}
a) Predictions of \textit{in vitro} free energies of nucleosome formation for a set of natural and
synthetic sequences measured using nucleosome dialysis (red circles) \cite{SI_Thastrom} and nucleosome exchange
(green circles). \cite{SI_Shrader_I,SI_Shrader_II}
Upper panel: DNABEND free energy predictions (with experimental error bars when available).
Lower panel: free energy predictions for the same dataset using a bioinformatics model
based on the alignment of sequences extracted from yeast mononucleosomes (the alignment model,
see Supplementary Methods). \cite{SI_Segal}
Artificial high affinity sequence 601 \cite{SI_Thastrom,SI_Thastrom_II} is highlighted in black. 
The free energies were computed using only the central 71 bp of the nucleosomal site:
$F = -\log \left( \sum_{i=1}^{L-L_{w}+1} \left[ e^{-E_i} + e^{-E^{(rc)}_i} \right] \right)$, where
$L$ is the length of each sequence,
$L_w = 71$ bp is the length of DNA bound by the $\mathrm{H3_{2}H4_{2}}$ tetramer (bp 39 through 109
in the 147 bp superhelix), and
$E_i$, $E^{(rc)}_i$ are the DNA elastic energies of bending the 71 bp long site into the superhelical shape
at sequence positions $i \dots i+70$ for the forward strand ($E_i$), and $i+70 \dots i$ for the reverse strand ($E^{(rc)}_i$).
In the alignment model, log-scores with genomic background frequencies were used instead of energies
($L^{g}$, cf. Equation (\ref{LogScore})).
b) Histograms of DNA elastic energies/alignment model log scores for the mouse genome sequences selected for their
ability to position nucleosomes (red) \cite{SI_Kubista_I}, or to impair nucleosome formation (blue). \cite{SI_Kubista_II}
Upper panel: DNABEND (with $w = 0.5$), lower panel: alignment model with genomic background
frequencies. \cite{SI_Segal} 
In sequences longer than 71 bp the most favorable energy/log score was used,
taking both forward and reverse strands into account.
We assume that in all nucleosome reconstitution experiments described above the selective pressure
was exerted mainly on the central stretch of the nucleosomal DNA
which interacts with the $\mathrm{H3_{2}H4_{2}}$ tetramer. \cite{SI_Thastrom_II}
}
\end{figure}

\begin{figure}[tbph]
\centering
\includegraphics[scale=0.7]{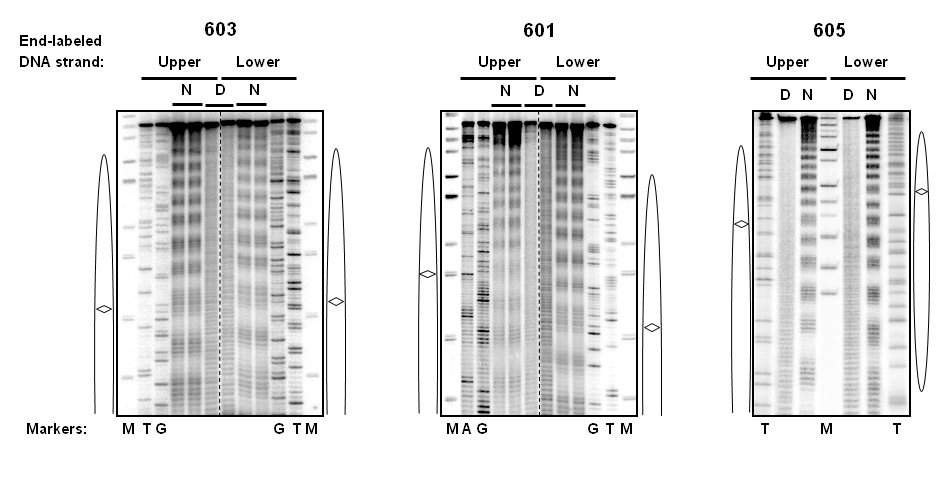}
\caption{
\textbf{Hydroxyl radical footprinting of the 601, 603 and 605 nucleosomal templates.}
Single-end-labeled histone-free DNA (D) or nucleosomes (N) were exposed to hydroxyl radicals,
DNA was purified and analyzed by denaturing PAGE (representative gels are shown).
The lower and upper DNA strands are shown in Supplementary Fig. 14. Nucleosome positions are indicated by ovals;
the positions of nucleosomal dyads in the gels are indicated by diamonds.
The A, T and G markers were obtained by primer extension using Taq DNA polymerase and terminating dideoxynucleotide
triphosphates (ddATP (A), ddTTP (T), and ddGTP (G), respectively).
M: end-labeled pBR322-MspI digest (DNA size markers).
}
\end{figure}

\begin{figure}[tbph]
\centering
\includegraphics[scale=0.7]{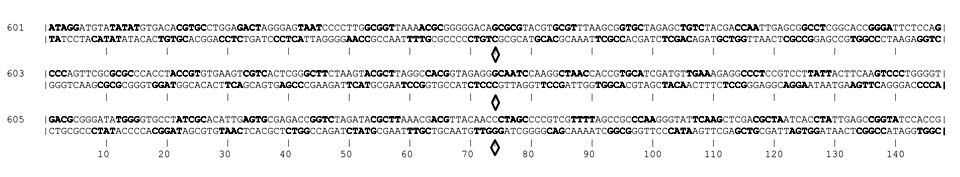}
\caption{
\textbf{Schematic description of the results of hydroxyl radical footprinting of the 601, 603 and 605 nucleosomal templates.}
The sites accessible to and protected from hydroxyl radicals are shown in bold and regular letters, respectively.
The accessible and protected nucleotides were identified by quantitative analysis of hydroxyl radical
footprinting patterns (Supplementary Fig. 13). The positions of nucleosomal dyads are indicated by diamonds.
}
\end{figure}

\begin{figure}[tbph]
\centering
\includegraphics[scale=0.85]{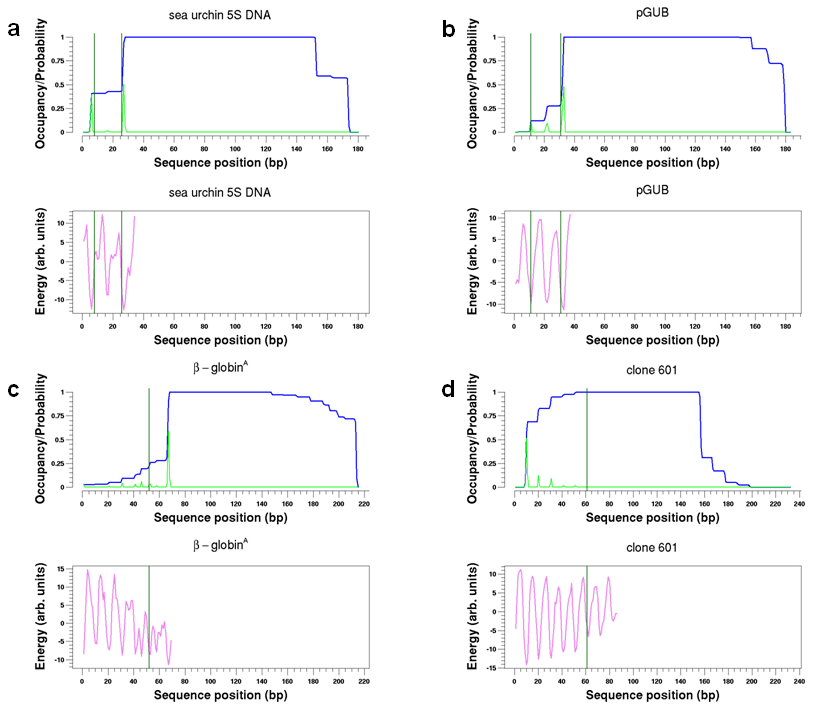}
\end{figure}

\newpage
\begin{figure}[!tbph]
\centering
\includegraphics[scale=0.85]{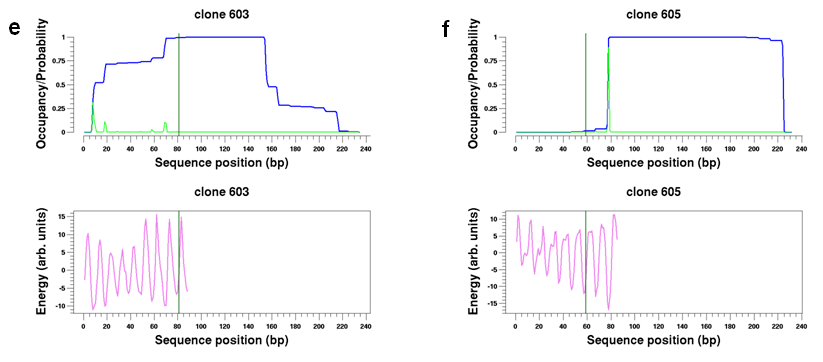}
\caption{
\textbf{DNABEND predictions of {\it in vitro} nucleosome positions.}
Probability of a nucleosome to start at each base pair (green), nucleosome occupancy (blue), and nucleosome formation energy (violet).
Vertical lines: experimentally known nucleosome starting positions, with bp coordinates listed in brackets below.
(a) The 180 bp sequence from the sea urchin 5S rRNA gene [bps 8,26]. \cite{SI_Flaus}
(b) The 183 bp sequence from the pGUB plasmid [bps 11,31]. \cite{SI_Kassabov}
(c) The 215 bp fragment from the sequence of the chicken $\beta-\mathrm{globin}^{\mathrm{A}}$ gene [bp 52]. \cite{SI_Davey}
(d,e,f) Synthetic high-affinity sequences \cite{SI_Lowary} 601 [bp 61], 603 [bp 81], and 605 [bp 59].
Experimentally known nucleosome starting positions are listed in a one-based coordinate system with consecutively numbered base pairs.
Nucleosomes on sequences 601, 603, and 605 were mapped by hydroxyl radical footprinting (Supplementary Figs. 13 and 14).
All DNA sequences used in this calculation are available on the Nucleosome Explorer website: \textit{http://nucleosome.rockefeller.edu}.
}
\end{figure}

\begin{figure}[tbph]
\centering
\includegraphics[scale=0.9]{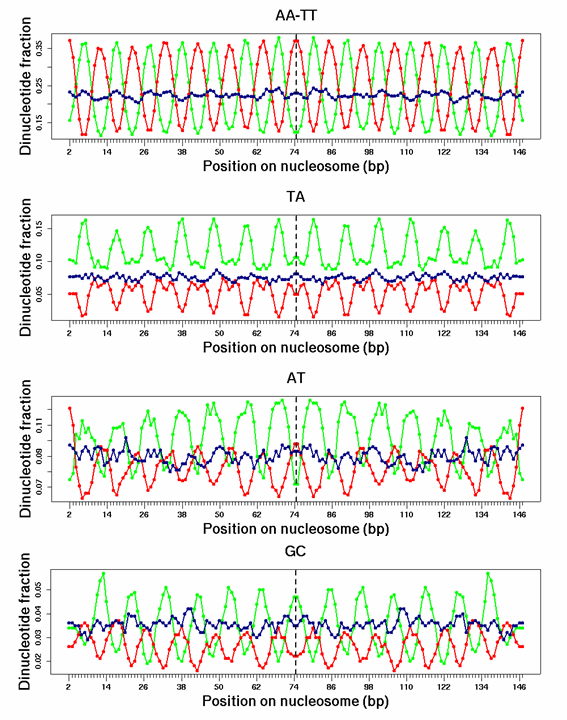}
\caption{
\textbf{DNABEND-selected sequences exhibit periodic dinucleotide patterns.}
Relative fractions of AA/TT, TA, AT, and GC dinucleotides
in the alignment of 1000 147 bp sequences from 
the \textit{S.cerevisiae} chromosome IV.
Green: sequences with lowest DNA elastic energies,
red: sequences with highest DNA elastic energies, dark blue: randomly picked sequences.
DNA elastic energies were predicted by DNABEND.
}
\end{figure}

\begin{figure}[tbph]
\centering
\includegraphics[scale=0.77]{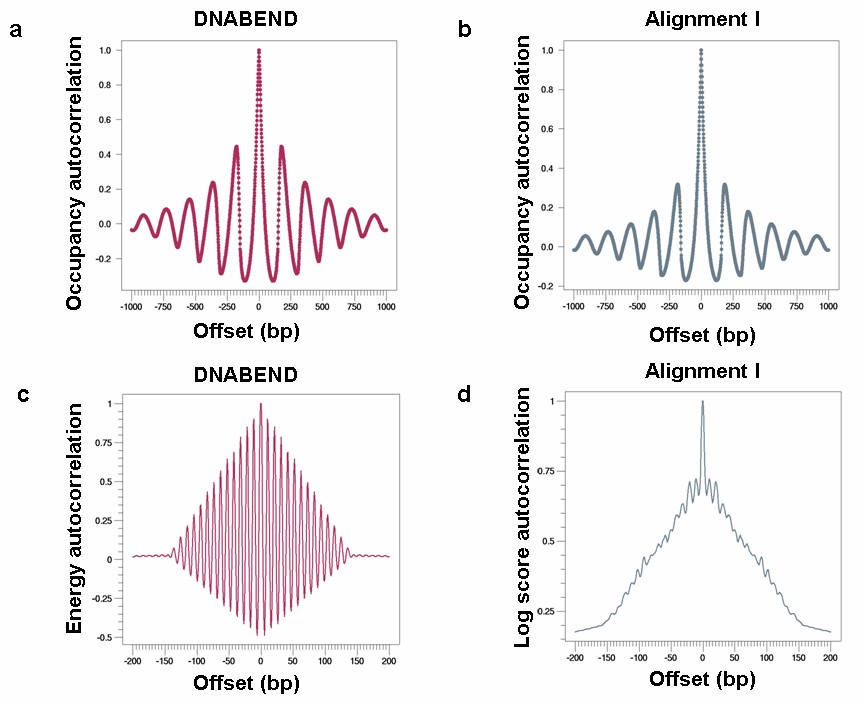}
\caption{
\textbf{Autocorrelation functions show periodicity in nucleosome occupancies, energies, and log scores.}
Autocorrelation function for nucleosome occupancies computed using DNABEND (a)
and alignment model I (b). \cite{SI_Segal}
Autocorrelation function for nucleosome energies computed using DNABEND (c),
and for log scores computed using alignment model I, Equation (\ref{LogScore:combined}) (d). \cite{SI_Segal}
Nucleosome occupancies are periodic due to steric exclusion, while energies and log scores are periodic
due to DNA helical twist.
}
\end{figure}

\begin{figure}[tbph]
\centering
\includegraphics[scale=0.77]{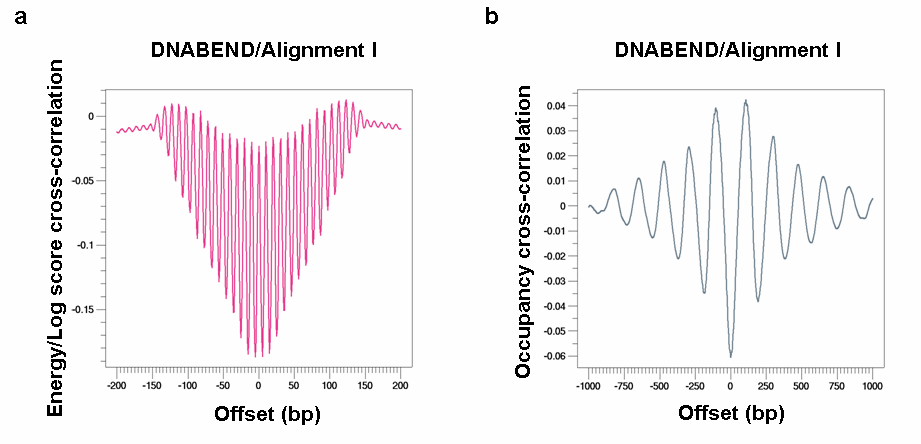}
\caption{
\textbf{Cross-correlation functions reveal low correlation between DNABEND and alignment model predictions.}
Cross-correlation functions between energies/log scores (a)
and occupancies (b) computed using DNABEND and alignment model I.
Both correlations are negative when computed at zero offset.
}
\end{figure}

\begin{figure}[tbph]
\centering
\includegraphics[scale=1.0]{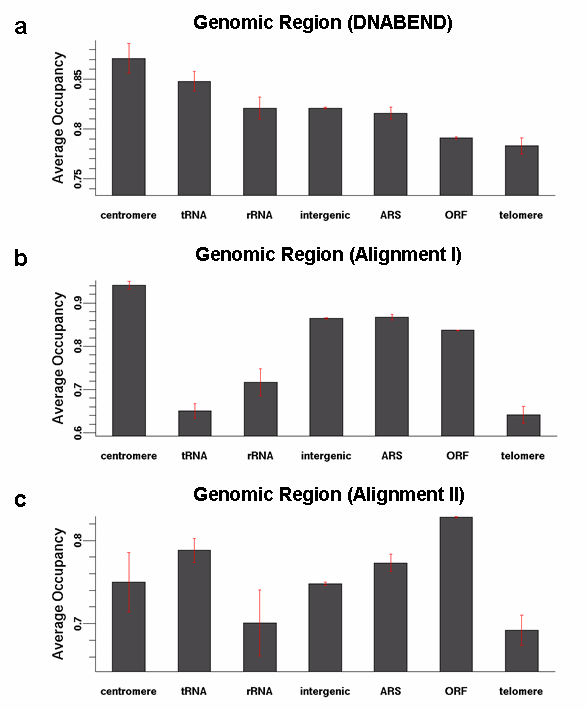}
\caption{
\textbf{Average nucleosome occupancies and standard errors for \textit{S.cerevisiae} genomic regions.}
(a) - DNABEND, (b) - alignment model I, \cite{SI_Segal} (c) - alignment model II.
}
\end{figure}

\begin{figure}[tbph]
\centering
\includegraphics[scale=0.9]{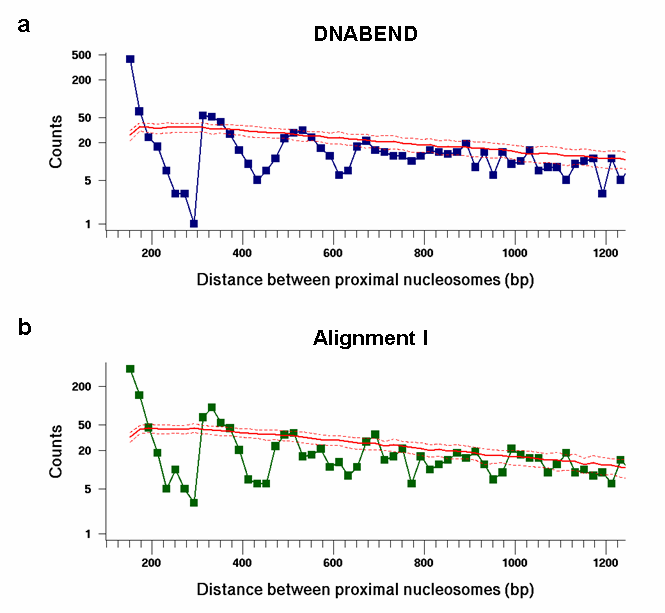}
\caption{
\textbf{Stable nucleosomes are arranged in a regular array induced by the periodic occupancy profile.}
Center-to-center distance counts for proximal stable nucleosomes ($P \ge 0.5$) predicted
using DNABEND (a) and alignment model I (b).
Red: null model in which the same number of stable nucleosomes is randomly positioned on the
genome without overlap. Solid red lines are mean values for 100 random placements, dashed red lines
are one standard deviation away. Alignment model II exhibits similar oscillations (data not shown).
}
\end{figure}

\begin{figure}[tbph]
\centering
\includegraphics[scale=0.9]{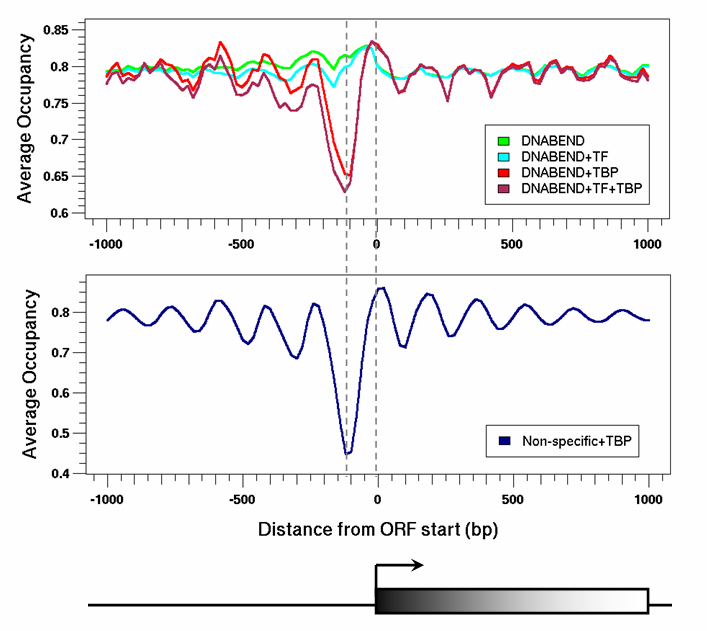}
\caption{
\textbf{Average nucleosome occupancy plotted with respect to translation start sites.}
Green - DNABEND nucleosomes only (same as in Fig. 3a),
cyan - DNABEND nucleosomes competing with TFs allowed to bind to functional sites, \cite{SI_MacIsaac_I}
red - DNABEND nucleosomes competing with TBPs (same as Fig. 3a),
maroon - DNABEND nucleosomes competing with both TFs and TBPs.
Dark blue - a non-specific nucleosome model (all nucleosome positions assigned a constant
energy) competing with TBPs.
}
\end{figure}

\begin{figure}[tbph]
\centering
\includegraphics[scale=0.9]{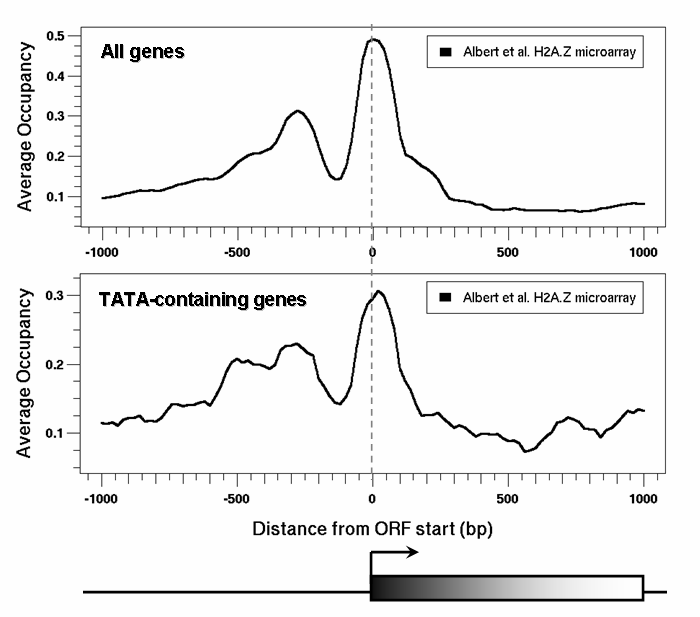}
\caption{
\textbf{Average H2A.Z nucleosome occupancy plotted with respect to translation start sites.}
The occupancy profile was derived as described in the Supplementary Methods,
starting from H2A.Z nucleosome positions determined
with high throughput DNA sequencing by Albert \textit{et al.} \cite{SI_Albert}
Note that nucleosome occupancy profiles were constructed by giving an approximately equal weight to each gene,
regardless of the total number of sequence reads in its vicinity. This procedure results in
occupancy profiles that are smoother than those based on weighting all sequence reads equally,
as shown in Albert \textit{et al.} \cite{SI_Albert}
}
\end{figure}

\begin{figure}[tbph]
\centering
\includegraphics[scale=0.9]{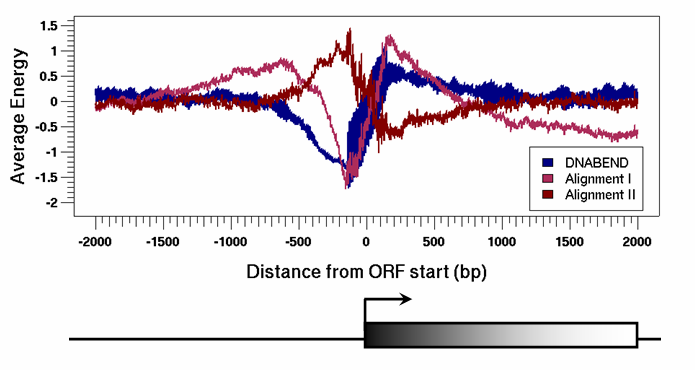}
\caption{
\textbf{Average nucleosome energies and log scores plotted with respect to translation start sites.}
Dark blue: DNABEND, maroon: alignment model I, dark red: alignment model II.
Nucleosome stability in the upstream regions is higher than average
with DNABEND and alignment model I, and lower than average with alignment model II.
}
\end{figure}

\begin{figure}[tbph]
\centering
\includegraphics[scale=0.95]{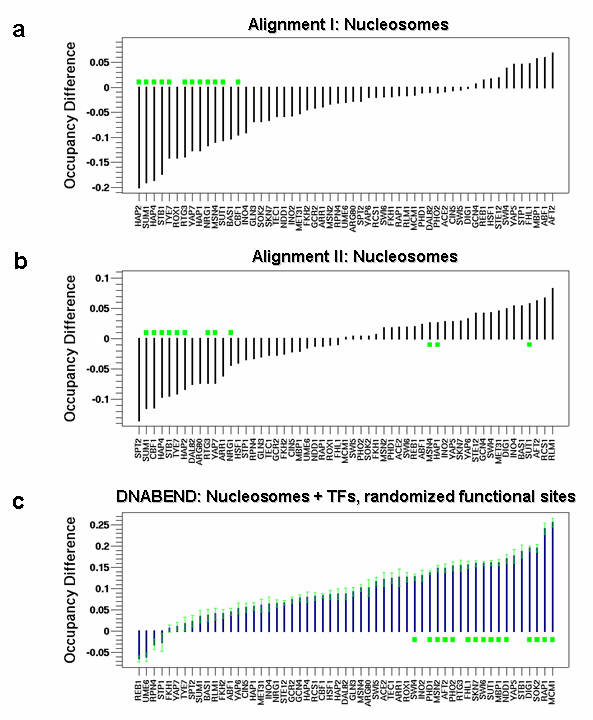}
\end{figure}

\begin{figure}[tbph]
\centering
\caption{
\textbf{Difference in the average nucleosome occupancy of functional and non-functional TF binding sites.} \cite{SI_MacIsaac_I}
The difference is negative if the functional sites are less covered by nucleosomes.
Green squares indicate statistically significant occupancy differences ($P < 0.05$).
(a) Nucleosomes only, alignment model I, \cite{SI_Segal}
(b) nucleosomes only, alignment model II,
(c) nucleosomes competing with TFs.
In (c), non-functional sites are the same as in Fig. 4a,
whereas the positions of functional sites have been randomized within intergenic regions
(we retained TF binding energy distributions of functional sites after randomization,
disregarding actual TF binding specificities for the purposes of this test).
Error bars correspond to 20 different realizations of functional site randomization.
Note that with DNABEND functional binding sites are covered by nucleosomes if their
positions are randomized, whereas according to alignment models functional binding
sites are intrinsically uncovered. As in Fig. 4a, we only consider TFs for which at least
20 binding sites are available in each (functional or non-functional) set.
}
\end{figure}

\begin{figure}[tbph]
\centering
\includegraphics[scale=0.92]{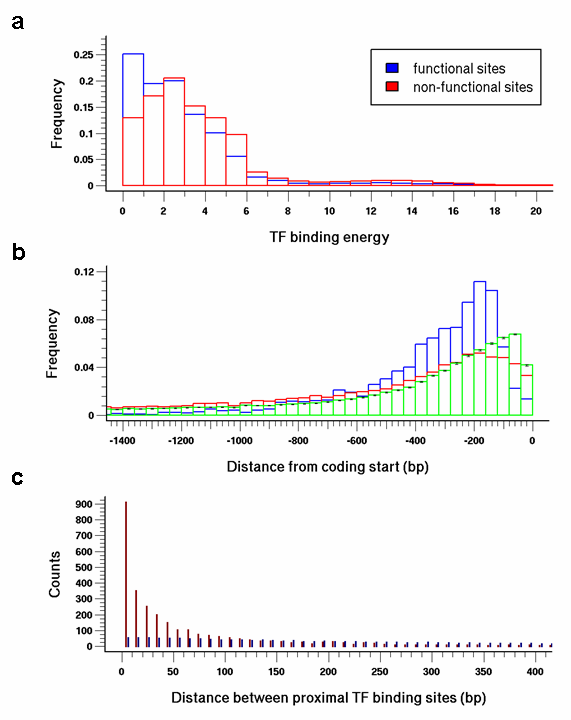}
\caption{
\textbf{Spatial distribution and binding energies of functional and non-functional TF binding sites.}
(a) Histogram of TF binding energies in functional (blue) and non-functional (red) sites. \cite{SI_MacIsaac_I}
Zero energy corresponds to the most favorable (consensus) sequence.
(b) Histogram of TF binding site positions with respect to the translation start site.
Blue - functional sites, red - non-functional sites, green - randomized functional sites.
(c) Histogram of the center-to-center distances between proximal TF sites shows TF site clustering.
Dark red - functional sites, dark blue - randomized functional sites.
Error bars are too small to be shown in the latter graph.
Randomized functional sites are as in Supplementary Fig. 24.
}
\end{figure}

\clearpage

\begin{figure}[tbph]
\centering
\caption{
\textbf{Predictions of nucleosome positions known from the literature: specific genomic loci.}
Orange ovals: experimentally mapped nucleosomes.
Upper panel: DNABEND-predicted nucleosome occupancy,
both with (maroon) and without (blue) other DNA-binding factors.
Lower panel: TBP occupancy (green) and TF occupancy (dark blue, dark orange, pink).
Black arrows indicate translation start sites.
To compute DNA-binding energies of TFs we scanned intergenic regions using PWMs
from Morozov \textit{et al.} \cite{SI_Morozov_II} or, if those were not available,
from MacIsaac \textit{et al.} \cite{SI_MacIsaac_I} (a few PWMs had a different origin as discussed below in
individual Figure captions).
TBP PWM was derived using an alignment of TATA box sites from Basehoar \textit{et al.}; \cite{SI_Basehoar}
since, unlike TFs, there were too many false positives we further restricted regions where TBP could bind using information from
the literature, and omitted TBP binding altogether if no such regions could be found.
The TF binding energy to the consensus (lowest energy) sequence was typically set to -9.0 kcal/mol, whereas for TBP it was always
-6.0 kcal/mol (cf. section \ref{TF:TBP} of Supplementary Methods).
Finally, we corrected obvious discrepancies
between measured and predicted nucleosome occupancies by changing the value of the chemical potential
(the concentration of the unbound histone octamer) in several cases, notably for $HML\alpha$, $HMR\mathbf{a}$ and
recombination enhancer loci,
and by omitting the 5bp linkers from the nucleosome model.
These changes affected only maroon curves. All predictions shown here are summarized in Figure 5 from the main text.
}
\end{figure}

\clearpage

\addtocounter{figure}{-1} 

\begin{subfigures}

\begin{figure}[tbph]
\centering
\includegraphics[scale=0.9]{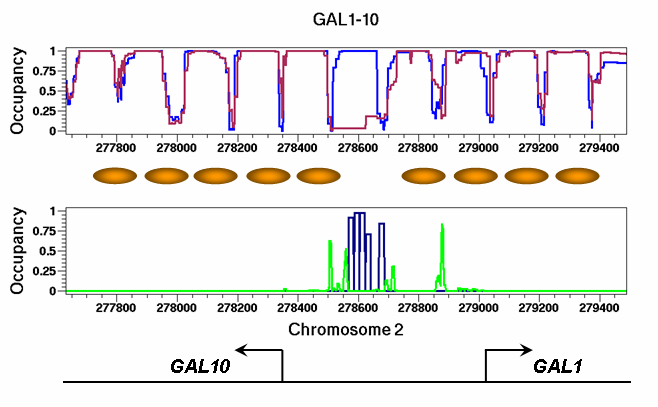}
\caption{
Detailed view of the \textit{GAL1-10} locus, with experimental nucleosome positions mapped by Li \textit{et al.} \cite{SI_Li}
Green: TBP, dark blue: GAL4. Several GAL4 factors bind closely spaced sites and
act cooperatively to displace a single nucleosome. Once the nucleosome is removed TBP is free to bind TATA sequences
in the nucleosome-depleted region.
}
\end{figure}

\begin{figure}[tbph]
\centering
\includegraphics[scale=0.9]{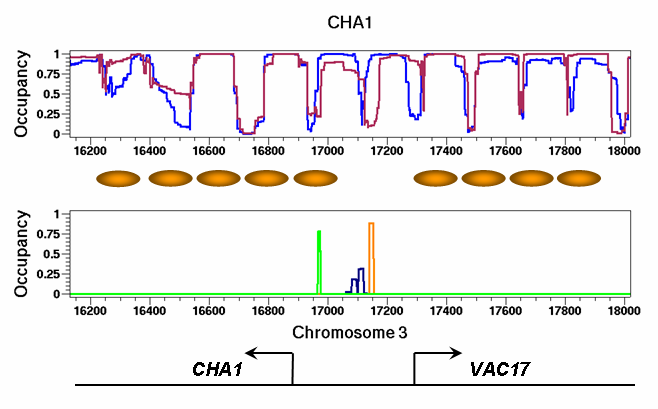}
\caption{
Detailed view of the \textit{CHA1} locus, with experimental nucleosome positions mapped by Moreira \textit{et al.} \cite{SI_Moreira}
Green: TBP, dark blue: CHA4, dark orange: ABF1.
}
\end{figure}

\begin{figure}[tbph]
\centering
\includegraphics[scale=0.9]{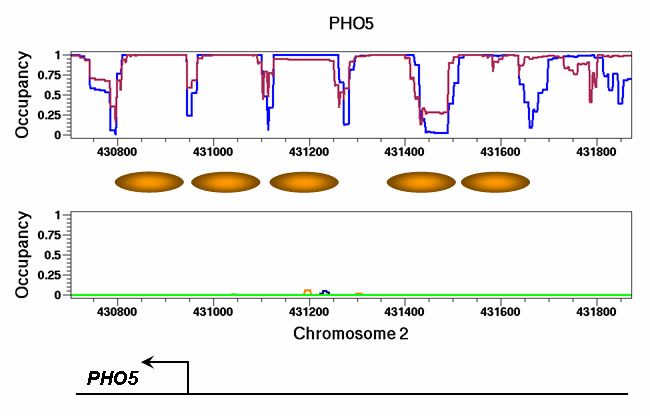}
\caption{
Detailed view of the \textit{PHO5} locus, with experimental nucleosome positions
mapped by H\"{o}rz \textit{et al.} \cite{SI_Almer,SI_Svaren_II,SI_Venter}
Green: TBP, dark blue: PHO2, dark orange: PHO4.
}
\end{figure}

\begin{figure}[tbph]
\centering
\includegraphics[scale=0.9]{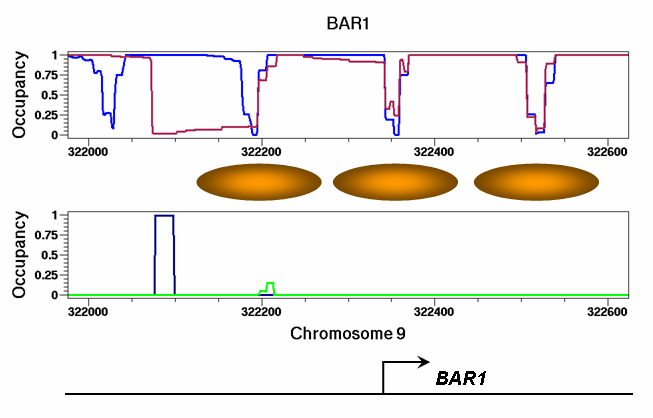}
\caption{
Detailed view of the \textit{BAR1} locus, with experimental nucleosome positions mapped by Shimizu \textit{et al.} \cite{SI_Shimizu}
Green: TBP, dark blue: $\mathrm{MAT\alpha2-MCM1-MAT\alpha2}$ (with PWM constructed using a combination of a TRANSFAC MCM1 PWM and a
structure-based 1mnm PWM \cite{SI_Morozov_II}). Note that the experimentally mapped nucleosomes are phased off the region occupied
by $\mathrm{MAT\alpha2-MCM1-MAT\alpha2}$; however, in the DNABEND simulation nucleosomes move to the left instead,
perhaps because we overestimate the stability of the nucleosome starting at bp 322200.
}
\end{figure}

\begin{figure}[tbph]
\centering
\includegraphics[scale=0.9]{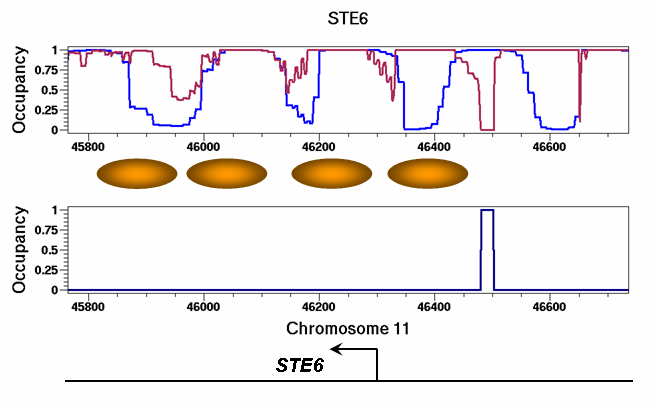}
\caption{
Detailed view of the \textit{STE6} locus, with experimental nucleosome positions mapped by Shimizu \textit{et al.} \cite{SI_Shimizu}
Dark blue: $\mathrm{MAT\alpha2-MCM1-MAT\alpha2}$ (same PWM as in Supplementary Fig. 26d).
Note that in contrast to the \textit{BAR1} locus (Supplementary Fig. 26d), nucleosomes are shifted in the right direction when
$\mathrm{MAT\alpha2-MCM1-MAT\alpha2}$ binds, phasing off the TF-occupied region.
}
\end{figure}

\begin{figure}[tbph]
\centering
\includegraphics[scale=0.9]{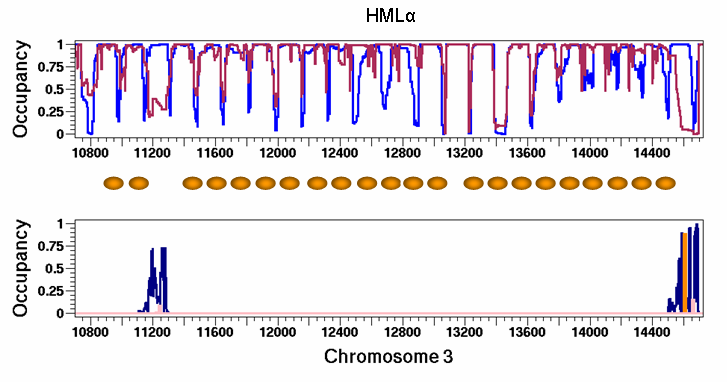}
\caption{
Detailed view of the $HML\alpha$ locus, with experimental nucleosome positions mapped by Weiss \textit{et al.} \cite{SI_Weiss_II}
Dark blue: origin recognition complex (PWM based on the consensus sequence from Ref.~\cite{SI_Bolon}),
dark orange: ABF1, pink: RAP1.
Note that two prominent gaps in the otherwise regularly spaced array of experimentally mapped nucleosomes
coincide with the regions of high TF occupancy.
}
\end{figure}

\begin{figure}[tbph]
\centering
\includegraphics[scale=0.9]{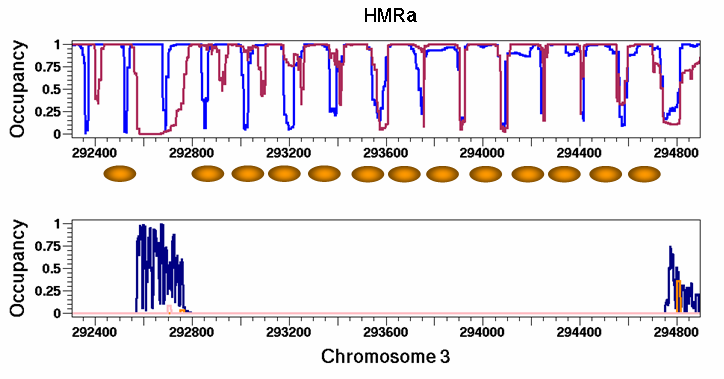}
\caption{
Detailed view of the $HMR\mathbf{a}$ locus, with experimental nucleosome positions mapped by Ravindra \textit{et al.} \cite{SI_Ravindra}
Dark blue: origin recognition complex (PWM based on the consensus sequence from Ref.~\cite{SI_Bolon}),
dark orange: ABF1, pink: RAP1. Similar to the $HML\alpha$ locus (Supplementary Fig. 26f), 
two prominent gaps in the otherwise regularly spaced array of experimentally mapped nucleosomes
coincide with the regions of high TF occupancy.
}
\end{figure}

\begin{figure}[tbph]
\centering
\includegraphics[scale=0.9]{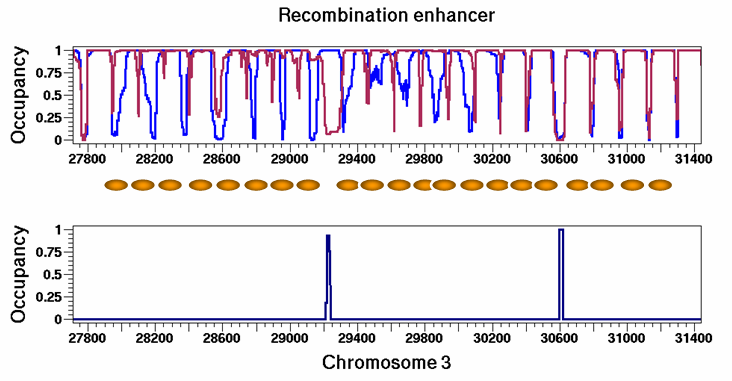}
\caption{
Detailed view of the recombination enhancer locus, with experimental nucleosome positions mapped by Weiss \textit{et al.} \cite{SI_Weiss_I}
Dark blue: $\mathrm{MAT\alpha2-MCM1-MAT\alpha2}$ (same PWM as in Supplementary Fig. 26d).
Similar to the $HML\alpha$ and $HMR\mathbf{a}$ loci (Supplementary Figs. 26f,g),
two prominent gaps in the otherwise regularly spaced array of experimentally mapped nucleosomes
coincide with the regions of high TF occupancy.
}
\end{figure}

\begin{figure}[tbph]
\centering
\includegraphics[scale=0.9]{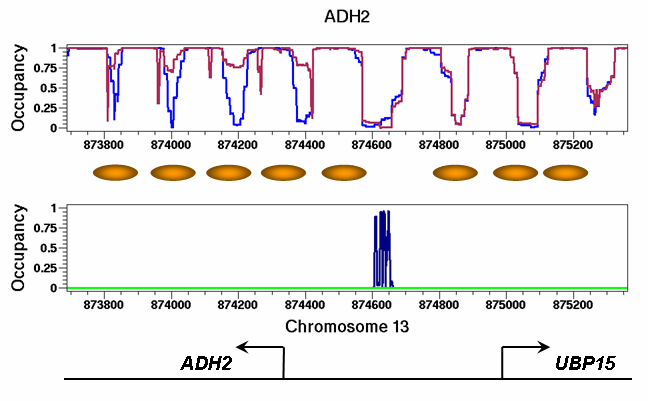}
\caption{
Detailed view of the \textit{ADH2-UBP15} locus, with experimental nucleosome positions
mapped by Verdone \textit{et al.} \cite{SI_Verdone}
Green: TBP, dark blue: ADR1.
}
\end{figure}

\begin{figure}[tbph]
\centering
\includegraphics[scale=0.9]{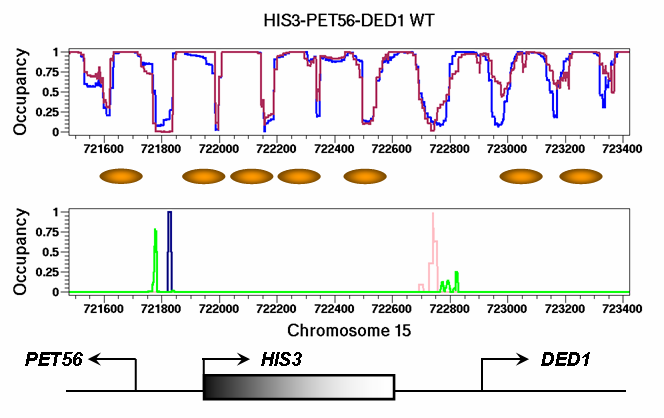}
\caption{
Detailed view of the \textit{HIS3-PET56-DED1} wildtype locus.
Upper panel: DNABEND-predicted nucleosome occupancy,
both with (maroon) and without (blue) other DNA-binding factors.
Lower panel: TBP occupancy (green), GCN4 occupancy (dark blue), and ABF1 occupancy (pink).
Orange ovals: experimental nucleosome positions mapped by Sekinger \textit{et al.} \cite{SI_Sekinger}
Black arrows indicate translation start sites.
We used a GCN4 weight matrix from Morozov \textit{et al.} \cite{SI_Morozov_II} and an ABF1 weight matrix
from MacIsaac \textit{et al.} \cite{SI_MacIsaac_I} to compute TF DNA-binding energies.
The TBP weight matrix was derived using an alignment of TATA box sites from Basehoar \textit{et al.} \cite{SI_Basehoar}
Note that the region around bp 721800 is intrinsically devoid of nucleosomes, \cite{SI_Sekinger}
allowing TFs to bind in the nucleosome-free gap.
}
\end{figure}

\begin{figure}[tbph]
\centering
\includegraphics[scale=0.9]{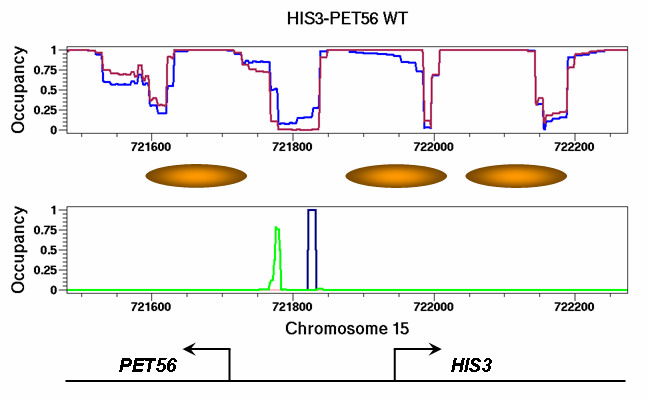}
\caption{
Detailed view of the \textit{HIS3-PET56} wildtype locus (zoom-in of (j)),
with experimental nucleosome positions mapped by Sekinger \textit{et al.} \cite{SI_Sekinger}
Green: TBP, dark blue: GCN4, pink: ABF1.
Note that the region around bp 721800 is intrinsically devoid of nucleosomes, \cite{SI_Sekinger}
allowing TFs to bind in the nucleosome-free gap.
}
\end{figure}

\begin{figure}[tbph]
\centering
\includegraphics[scale=0.9]{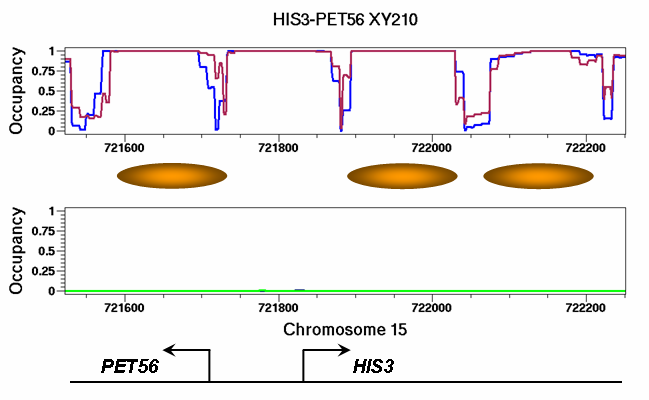}
\caption{
Detailed view of the \textit{HIS3-PET56} XY210 sequence deletion locus,
with experimental nucleosome positions mapped by Sekinger \textit{et al.} \cite{SI_Sekinger}
Green: TBP, dark blue: GCN4.
Note that DNABEND likely overpredicts the stability of the nucleosome centered at bp 721800.
As a result, this nucleosome is not displaced in competition with TFs (whose binding energies
may also be underestimated).
}
\end{figure}

\begin{figure}[tbph]
\centering
\includegraphics[scale=0.9]{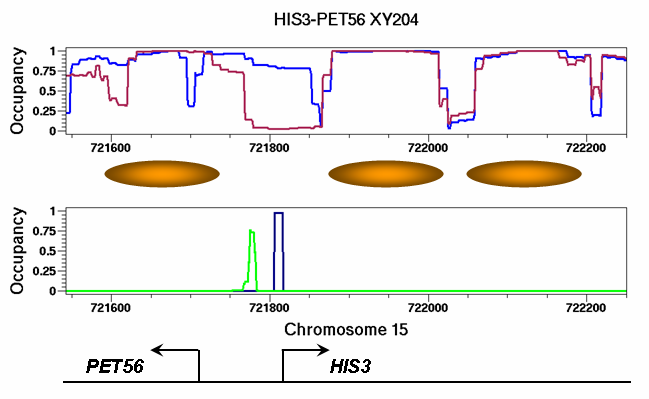}
\caption{
Detailed view of the \textit{HIS3-PET56} XY204 sequence deletion locus,
with experimental nucleosome positions mapped by Sekinger \textit{et al.} \cite{SI_Sekinger}
Green: TBP, dark blue: GCN4. Note that unlike the case of the \textit{HIS3-PET56} wildtype locus (Supplementary Fig. 26k),
TF occupancy improves correlation with experiment by moving nucleosomes to the left of the TF-covered region.
}
\end{figure}

\begin{figure}[tbph]
\centering
\includegraphics[scale=0.9]{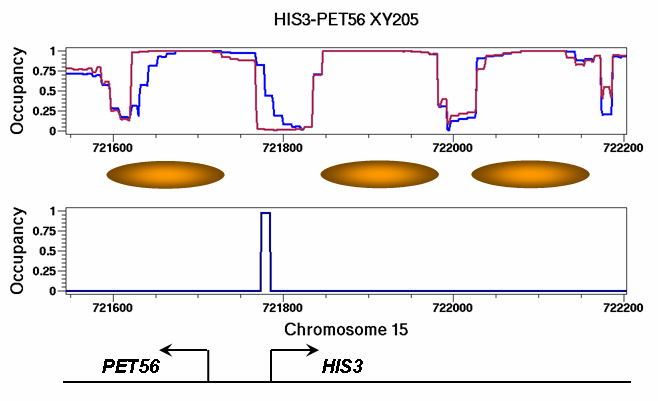}
\caption{
Detailed view of the \textit{HIS3-PET56} XY205 sequence deletion locus,
with experimental nucleosome positions mapped by Sekinger \textit{et al.} \cite{SI_Sekinger}
Dark blue: GCN4. Note that the intrinsic gap around bp 721800 is enlarged through the
action of GCN4, though its width is still insufficient compared to experiment.
}
\end{figure}

\end{subfigures}

\clearpage

\begin{figure}[tbph]
\centering
\includegraphics[scale=0.88]{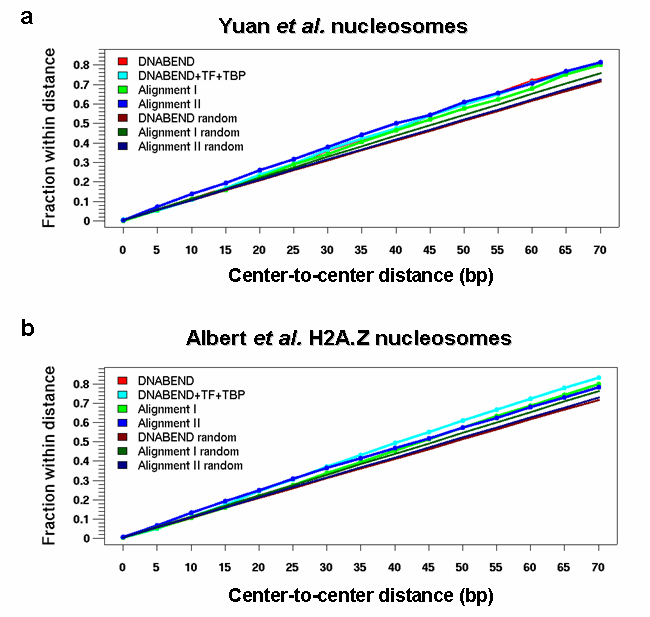}
\caption{
\textbf{Predictions of nucleosome positions mapped in genome-wide assays.}
Fraction of correctly predicted nucleosomes as a function of the minimal center-to-center distance between predicted
nucleosomes and (a) nucleosomes mapped using Yuan \textit{et al.} microarray assay, \cite{SI_Yuan}
(b) H2A.Z nucleosomes mapped using Albert \textit{et al.} DNA sequencing approach. \cite{SI_Albert}
In the null (random) models, non-overlapping nucleosomes are placed on the genome with probability $P = 1.0$. 
Distances between neighboring nucleosomes
are sampled from a Gaussian distribution with the mean chosen to reproduce the average occupancy predicted by the
actual model, and the standard deviation set to 0.5 of the mean (negative linker lengths are not allowed).
Thus in the null models nucleosomes are positioned non-specifically but on average form a regular array.
The only parameter borrowed from the actual model is the predicted average occupancy.
The center-to-center nucleosome distance is determined by locating in the periodic occupancy profile
the nearest region with occupancy of at least 0.35 over the nucleosomal length.
Error bars that correspond to 100 random realizations of the null models are too small to be shown.
Note that in both (a) and (b) red and cyan curves nearly overlap; adding TF and TBP binding to the alignment models does not
improve their performance (data not shown).
}
\end{figure}

\clearpage

\newpage


\end{document}